\documentclass[a4paper,11pt,nofootinbib]{revtex4}
\usepackage{graphicx,amssymb,bm,latexsym,color,epsf}
\makeatletter
\@addtoreset{equation}{section}

\makeatother
\newcommand{\be}{\begin{equation}}
\newcommand{\ee}{\end{equation}}
\newcommand{\bea}{\begin{eqnarray}}
\newcommand{\eea}{\end{eqnarray}}

\newcommand{\bR}{\bar R}

\newcommand{\bnabla}{\bar\nabla}

\newcommand{\bg}{\bar g}

\newcommand{\tilh}{{h^T}}
\newcommand{\tth}{{h^{TT}}}

\newcommand{\bphi}{\bar\phi}
\newcommand{\tr}{\mathrm{tr}}

\newcommand{\Lie}{{\cal L}}

\newcommand{\tG}{\tilde G}
\newcommand{\tL}{\tilde \Lambda}
\begin{document}

\begin{titlepage}

\title{Search of scaling solutions in scalar-tensor gravity}

\author{Roberto Percacci}
\email{percacci@sissa.it}
\affiliation{
SISSA, via Bonomea 265, I-34136 Trieste
\\
and 
INFN, Sezione di Trieste, Italy
}
\author{Gian Paolo Vacca}
\email{vacca@bo.infn.it}
\affiliation{INFN, Sezione di Bologna,
via Irnerio 46, I-40126 Bologna}

\begin{abstract}
We write new functional renormalization group equations for a scalar nonminimally coupled to gravity.
Thanks to the choice of the parametrization and of the gauge fixing they are simpler than older equations and avoid some of the difficulties that were previously present.
In three dimensions these equations admit, at least for sufficiently small fields, a solution that
may be interpreted as a gravitationally dressed Wilson-Fisher fixed point.
We also find for any dimension $d>2$ additional analytic scaling solutions which we study for 
$d=3$ and $d=4$. One of them corresponds to the fixed point of the Einstein-Hilbert truncation,
the others involve a nonvanishing minimal coupling.
\end{abstract} 

\maketitle

\end{titlepage}
\newpage
\setcounter{page}{2}

\section{Introduction}

In the quest of an UV-complete quantum field theory of gravity,
the search for a fixed point using functional renormalization
group methods has reached the point where one may hope to go beyond finitely
many couplings and study entire functional classes of truncations.
The best studied case is that of $f(R)$ actions, where a fixed point is known
to exist, and to exhibit nice stability properties, when $f$ is a polynomial \cite{cpr1,cpr2,macsau}.
The most advanced calculations have now reached order $R^{34}$ \cite{flnr,flr}.
However, the radius of convergence of the Taylor series of $f$ around the origin
is finite and there is not much to be gained by pushing the expansion much further.
Rather, one would like to find a scaling solution for the whole function $f$.
Several studies have shed light on various aspects of this issue but have
so far failed to reach a convincing conclusion, at least in four dimensions 
\cite{caravelli,dietz1,demmel1,dietz2,benedetti2,demmel2}.
An important fact that has been pointed out in \cite{dietz1} is that the equation
of \cite{cpr2,macsau} does not admit complete solutions. The simpler equation proposed in
\cite{caravelli} admits solutions at least for positive $R$ but then it was shown
in \cite{dietz2} that all perturbations around them are redundant, 
i.e. can be absorbed by field redefinitions.
One thus has to find a ``better'' equation, i.e. one admitting a discrete set of solutions
with non-redundant perturbations, or else show that no such equation exists.
In order to gain some understanding of what may be wrong with the equations
of \cite{cpr2,macsau,caravelli}, it has been shown in \cite{dietz3} that the use of background-dependent
regulators in the flow equation for a scalar field can artificially lead to similar pathologies.
It is therefore important to understand whether different ways of applying the
background field method could solve this issue.

In this paper we will discuss similar problems but in a different context, 
namely a scalar field non-minimally coupled to gravity.
We will consider Effective Average Actions (coarse-grained effective actions
depending on a cutoff $k$, usually abridged EAA) of the functional form:
\begin{equation}
\label{action}
\Gamma_k [\phi,g] = \int\,{d}^{d} x\,\sqrt{g}\left(V(\phi)-F(\phi)R
+\frac{1}{2}  g^{\mu \nu} \partial_{\mu} \phi \partial_{\nu} \phi \right)
+S_{GF}+S_{gh}\ ,
\end{equation}
where $S_{GF}$ and $S_{gh}$ are gauge-fixing and ghost terms.
The usual Einstein-Hilbert action is contained in this truncation 
as the constant ($\phi$-independent) part of the action,
while switching off gravity (i.e. setting $g_{\mu\nu}=\delta_{\mu\nu}$)
reduces the system to the well-studied
Local Potential Approximation (LPA) of the scalar field.

There are several good reasons to study such actions.
On one hand they may have direct applications to cosmology \cite{hprw}.
At the classical level, they are related via some field redefinitions
to the $f(R)$ class of actions. 
\footnote{This has motivated the study of the RG flow
of Brans-Dicke-type actions in \cite{guarnieri}.}
Whether the classical equivalence can be preserved at the quantum level is doubtful at present,
but not completely settled.
From the point of view of this work, they have the important advantage that
the scalar and gravity subsectors in isolation are well-understood.
In particular, in $d=3$ the LPA admits a solutions that is a good approximation
to the Wilson-Fisher fixed point \cite{morris3,LZ}
and pure gravity admits a fixed point that is relatively weakly coupled,
independent of the presence or absence of higher derivative terms
\cite{lauschereuter,ps,ppps,ohta,pohta}.
These two facts lead us to suspect that the coupled system also should have a fixed point.

An early study of one-loop divergences of the form (\ref{action}),
in fact including also a prefactor $Z(\phi)$ for the kinetic term,
was made in \cite{bkk}, see also \cite{steinwachs}.
Here we shall derive the RG flow of $V_k$ and $F_k$
from the functional RG equation~\cite{wett1,morris1}, 
\be
\label{erge}
\dot \Gamma_k [\Phi]=\frac{1}{2}{\rm STr}\left[ \left(\Gamma^{(2)}[\Phi]+R_k\right)^{-1} \dot R_k \right]\ ,
\ee
where the dot stands for the partial derivative with respect to $t=\log k/k_0$,
and $R_k$ is the operator which realises the coarse-graining procedure.
The flow of $V_k$ and $F_k$ is a system of coupled PDE's
while the fixed-point equations form a system of coupled ODE's.
Solving such equations will be the main challenge of this work.

The flow equations for the theory (\ref{action}) had been derived 
earlier in \cite{narain}.
Fixed point solutions had been found for polynomial truncations but they do not
have the desired properties, as we shall recall in some detail below.
It is possible that, just as in the $f(R)$ case, 
this is due to the way the exact equation has been approximated,
and in particular to the implementation of the background field method.
Therefore we derive alternative equations based on a different
definition of the quantum-background split and a different gauge choice.
We will then see that the new equations admit nontrivial scaling solutions.

This paper is organized as follows. In section II we 
discuss the issues encountered by the old flow equations.
In section III we motivate the use of the exponential parametrization for the metric
and of ``physical'' gauge choices, in particular of the ``unimodular'' gauge.
We then derive new flow equations for $F$ and $V$, valid in any dimension.
For simplicity we discuss first the case when the terms proportional to $\dot F$
in the r.h.s. of the equations are neglected.
In sections IV and V we discuss in some detail the solutions of these equations in $d=3$ and $4$
respectively.
In section VI we briefly describe some results when the terms proportional to $\dot F$
are retained. 
Section VII contains our conclusions.
Three appendices contain a discussion of a multiplicative background field method,
of functional Jacobians and some results in arbitrary dimensions.

\section{Old equations and their ailments.}

Flow equations for the functions $F$ and $V$ have been derived in \cite{narain},
see also \cite{hprw}.
They were then further simplified by Taylor expanding $F$ and $V$ around $\phi=0$
to some finite order, and fixed points have been searched within the resulting
finite dimensional theory space.
In $d=4$ the only nontrivial solution had constant $f$ and $v$.
It represents a non-interacting scalar field minimally coupled to
the well-known fixed point of pure gravity in the Einstein-Hilbert truncation.
The absence of other solutions was perhaps not too surprising, given that
such solutions do not exist for the pure scalar theory.
In $d=3$, however, pure scalar theory admits a nontrivial scaling solution,
the well-known Wilson-Fisher fixed point.
In the simplest approximation, known as the Local Potential Approximation (LPA),
(\ref{erge}) reduces to the following equation for the dimensionless potential $v(\varphi)$:
\be
\label{vdotscalar}
\dot v=
-3\,v+\frac{1}{2}\varphi\,v'+
\frac{1}{6\pi^2(1+v'')}
\ee
The solution to this equation can be obtained by a variety of semi-analytic
and numerical methods.
In view of this, there is perhaps greater reason to expect that
a nontrivial scaling solution may exist also for the system
of the scalar field coupled to gravity.
However, in \cite{narain} no such solution was found.
A Taylor expansion around $\phi=0$ yielded fixed points
all of whose Taylor coefficients are negative.
Even if this corresponded to a genuine fixed point, it would not be
physically acceptable.
If a fixed point existed and was analytic at $\phi=0$ it would show up as
a fixed point for the Taylor coefficients, so the failure to find a fixed point
for the latter implies that the functional equations do not have a 
global solution either.
The question then arises whether this reflects a genuine physical
property of the system, or some problem with the equations.

In order to discuss this we will not need to consider the whole
equations, it will suffice to look at one term that comes from
the contribution of the spin two excitations, namely
\be
\label{singular}
\dot v=\frac{1}{3\pi^2}
\left[\frac{f}{f-v}+\ldots\right]
\ee
In a polynomial expansion of the solution around $\varphi=0$, 
it turns out that $v(0)<f(0)$ \cite{narain}.
On the other hand, for large $\varphi$ one expects the solutions to behave like
$v=A\varphi^6+\ldots$ and $f=B\varphi^2+\ldots$,
where the dots stand for inverse powers of $\varphi^2$.
A solution with these boundary conditions would have to cross the singularity at $v=f$. 
Although this cannot be ruled out, it is likely that the 
failure to find physically acceptable polynomial fixed points is related to
the existence of this singularity.

This conclusion is reinforced by the following two observations.
First, that this issue concerns the behavior of the dimensionless potential 
when the dimensionless field $\varphi=\phi \,k^{-(d-2)/2}$ becomes large.
For $d>2$ this is therefore an infrared issue.
Second, when $F$ and $V$ are constant, one can identify
\be
\label{newt}
V=\frac{2\Lambda}{16\pi G}\ ;
\qquad
F=\frac{1}{16\pi G}
\ee
and the fraction $f/(f\!-\!v)$ reduces to $1/(1\!-\!2\Lambda/k^2)$.
The singularity we are discussing is therefore a 
generalization of the well-known infrared singularity
at $\Lambda=k^2/2$ that appears in most treatments of the gravitational flow equation.

This singularity is an artifact of the way the beta functions for $v$ and $f$ are constructed.
The inverse propagator for the transverse, traceless spin-2 components $h_{\mu \nu}^T$ 
is given by:
\begin{equation}
\label{eq:sp2propgrav}
F\left(-\nabla^2+\frac{d^2-3d+4}{d(d-1)} \, R\right)-V \, . 
\end{equation}
The last term comes from the expansion of the $\sqrt{g}$ in the potential term,
which in the standard linear background field expansion contains
terms of the form $h_{\mu\nu}h^{\mu\nu}$.
The origin of the troublesome term in the flow equation for the potential is 
this propagator, with $R$ set equal to zero.
But if $V\not=0$, putting $R$ to zero means that we evaluate the flow equation
on a configuration that is far off shell.
Let us see what would happen if we evaluated the equation on shell.
For constant $\phi$ the trace of the equation of motion of the metric implies
\be
\label{treom}
F R=\frac{d}{d-2}V\ .
\ee
If we use this relation to eliminate the $R$ term, the spin-2 inverse propagator becomes
\be
F(-\bnabla^2)+\frac{2V}{(d-1)(d-2)}\ .
\ee%
This would contribute to the flow equation of $v$ a term
\be
\label{singular}
\dot v=\frac{1}{3\pi^2}
\left[\frac{f}{f+\frac{2}{(d-1)(d-2)}v}+\ldots\right]
\ee
where the troublesome singularity at $v=f$ is no longer present.
This is strong evidence that the singularity at $v=f$ is unphysical.

From the discussion above it is tempting to try and expand the
flow equation around a solution that is (nearly) on shell.
The virtues of such an approach have been discussed previously by
Benedetti \cite{benedetti_onshell} and Falls \cite{falls}.
In the following we will not pursue this idea,
but rather we will employ a different parametrization of the field
and choice of gauge fixing that automatically avoid the issue.

\section{The new flow equations}

\subsection{Exponential parametrization}

Instead of the traditional linear quantum-background split
$g_{\mu\nu}=\bar g_{\mu\nu}+h_{\mu\nu}$ we shall use in this paper
an exponential parametrization
\be
\label{decomp}
g_{\mu\nu}=\bg_{\mu\rho}(e^h)^\rho{}_\nu
\ee
where $\bar g$ is a fixed but arbitrary background.
This expansion has been used previously in \cite{kkn}. 
See also \cite{nink} for a recent discussion
in a context that is closer to the present one.
Some geometrical motivation for the use of this formula is given in Appendix A.
We assume in this paper that the path integral measure is simple
when expressed in terms of the field $h$ thus defined.
We discuss in appendix B the Jacobian relating this measure
to the one of the linear parametrization.

We will use the background metric $\bg$ to raise and lower indices.
Then due to the symmetry of $g_{\mu\nu}$ and $\bg_{\mu\nu}$ also the tensor 
$h_{\mu\nu}=\bg_{\mu\rho}h^\rho{}_\nu$ is symmetric.
We have
\bea
g_{\mu\nu}&=&\bg_{\mu\nu}+h_{\mu\nu}
+\frac{1}{2}h_{\mu\lambda}h^\lambda{}_\nu+\ldots
\\
g^{\mu\nu}&=&\bg^{\mu\nu}-h^{\mu\nu}
+\frac{1}{2}h^{\mu\lambda}h_\lambda{}^\nu+\ldots
\eea
In contrast to the usual linear split, here also the covariant metric is
nonpolynomial in the quantum field $h^\mu{}_\nu$.
Another significant difference is that,
due to the formula $\det e^h=e^{\tr h}$, only the trace part of $h$
enters in the definition of the determinant, at all orders.
As a result $\sqrt{g}$ does not contribute to the action of traceless fluctuations,
which are therefore independent of the potential.
We can split
\be
h^\mu{}_\nu=\tilh^\mu{}_\nu+2\omega\delta^\mu_\nu
\ee
where $\tr h=2d\omega$ and $\tilh$ is tracefree.
Then
\be
\sqrt{g}=e^{d\omega}\sqrt{\bg}=\sqrt{\bg}\left(1+d\omega+\frac{1}{2}d^2\omega^2+\ldots\right)\ .
\ee

For the scalar field we also expand around a background $\bar\phi$:
\be
\phi=\bar\phi+\delta \phi\ .
\ee
We then expand the action (\ref{action}) to second order in $h$ and $\delta\phi$.
Collecting all the terms we find
\bea
\int d^dx\sqrt{\bg}\Biggl[&&\!\!\!\!\!\!
F(\bphi)\Bigl(\frac{1}{4}h_{\mu\nu}(-\bnabla^2)h^{\mu\nu}
+\frac{1}{2}h_{\mu\nu}\bnabla^\mu\bnabla^\rho h_\rho{}^\nu
-\frac{1}{2}(\tr h)\bnabla_\mu\bnabla_\nu h^{\mu\nu}
+\frac{1}{4}(\tr h)\bnabla^2 (\tr h)
\nonumber\\
&&
-\frac{1}{2}\bR_{\mu\rho\nu\sigma}h^{\mu\nu}h^{\rho\sigma}
+\frac{1}{2}\bR_{\mu\nu}h^{\mu\nu}(\tr h)
-\frac{1}{8}\bR\,(\tr h)^2\Bigr)
\nonumber\\
&&
-F'(\bphi)\left(\bnabla_\mu\bnabla_\nu h^{\mu\nu}
-\bnabla^2(\tr h)-\bR_{\mu\nu}h^{\mu\nu}+\frac{1}{2}\bR\,(\tr h)\right)\delta\phi
\nonumber\\
&&
+\frac{1}{2}\delta\phi(-\bnabla^2+V''(\bphi)-F''(\bphi)\bR)\delta\phi
+\frac{1}{2}V'(\bphi)(\tr h)\delta\phi
+\frac{1}{8}V(\bphi)(\tr h)^2
\Biggr]
\eea

This is identical to equation (6) in \cite{narain},
which was derived using a linear split, except for two terms that are missing here:
\be
\label{missing}
-\frac{1}{2}F(\bphi)\bR^{\mu\nu}h_{\mu\rho}h^\rho{}_\nu
-\frac{1}{4}(V(\bphi)-F(\bphi)\bR)h_{\mu\nu}h^{\mu\nu}\ .
\ee 
The latter came from the expansion to second order of the square root of
the determinant of $g$. It is absent here because in the exponential
parametrization the determinant depends only on the trace part of $h$.

We then proceed with the York decomposition for the tracefree part of $h$:
\be
\label{york}
\tilh_{\mu\nu}=\tth_{\mu\nu}
+\bnabla_\mu\xi_\nu
+\bnabla_\nu\xi_\mu
+\bnabla_\mu\bnabla_\nu\sigma
-\frac{1}{d}\bg_{\mu\nu}\bnabla^2\sigma\ ,
\ee
where $\bar\nabla^\mu h^{TT}_{\mu\nu}=0$ and $\bar\nabla^\mu\xi_\mu=0$.
As usual it is convenient to further redefine
\be
\label{redef}
\xi'_\mu=\sqrt{-\bar\nabla^2-\frac{\bar R}{d}}\xi_\mu\ ;\qquad
\sigma'=\sqrt{-\bar\nabla^2}\sqrt{-\bar\nabla^2-\frac{\bar R}{d-1}}\sigma\ .
\ee

Collecting all terms we can rewrite the quadratic action in terms
of the independent fields $\tth$, $\xi'$, $\sigma'$, $\omega$ and $\delta\phi$:
\bea
&&\int dx\sqrt{\bg}\Biggl[
F(\bphi)\Biggl(
\frac{1}{4}\tth_{\mu\nu}\left(-\bnabla^2+\frac{2\bR}{d(d-1)}\right)\tth^{\mu\nu}
-\frac{(d-1)(d-2)}{4d^2}\sigma'\left(-\bnabla^2\right)\sigma'
\nonumber\\
&&
-\frac{(d-1)(d-2)}{d}\omega\sqrt{(-\bnabla^2)\left(-\bnabla^2-\frac{\bR}{d-1}\right)}\sigma'
-(d-1)(d-2)\omega\left(-\bnabla^2+\frac{(d-2)\bR}{2(d-1)}\right)\omega\Biggr)
\nonumber\\
&&
-F'(\bphi)\frac{d-1}{d}\delta\phi\left(
\sqrt{(-\bnabla^2)\left(-\bnabla^2-\frac{\bR}{d-1}\right)}\sigma'
+2d\left(-\bnabla^2+\frac{(d-2)\bR}{2(d-1)}\right)\omega
\right)
\nonumber\\
&&
+\frac{1}{2}\delta\phi(-\bnabla^2+V''(\bphi)-F''(\bphi)\bR)\delta\phi
+V'(\bphi)d\omega\delta\phi
+\frac{1}{2}V(\bphi)d^2\omega^2\Biggr]
\label{decomposed}
\eea
Note the absence of $\xi'$ from the expansion.
Also note that the kinetic operator of the $\omega$ field is not the
conformal scalar operator (which has a factor 4 instead of 2 in the
denominator).

\subsection{Gauge choice}

At this point we have to choose a gauge.
In order to simplify the equations as much as possible we will choose
a ``physical'' gauge, which amounts to putting the gauge-variant
components of $h_{\mu\nu}$ to zero.
Such gauges have been discussed earlier in a similar context in \cite{ppps},
see also \cite{gaber,zhang,mavro}.

The transformation of the metric under an infinitesimal diffeomorphism $\epsilon$
is given by the Lie derivative
\be
\label{transfg}
\delta_\epsilon g_{\mu\nu}=
\Lie_\epsilon g_{\mu\nu}
\equiv
\epsilon^\rho\partial_\rho g_{\mu\nu}
+g_{\mu\rho}\partial_\nu\epsilon^\rho
+g_{\nu\rho}\partial_\mu\epsilon^\rho\ .
\ee
As usual, we have to define transformations of $\bg$ and $h$ that,
used in (\ref{decomp}), yield (\ref{transfg}).
The simplest one is the background transformation.
If we treat $\bg$ and $h$ as tensors under $\delta_\epsilon$, i.e.
\be
\delta^{(B)}_\epsilon\bg_{\mu\nu}=\Lie_\epsilon \bg_{\mu\nu}\ ;
\qquad
\delta^{(B)}_\epsilon h^\mu{}_\nu=\Lie_\epsilon h^\mu{}_\nu\ .
\ee
then also
\be
\delta^{(B)}_\epsilon(e^h)^\mu{}_\nu=\Lie_\epsilon(e^h)^\mu{}_\nu
\ee
and (\ref{transfg}) follows.
By definition, the ``quantum'' gauge transformation of $h$ is such as to reproduce (\ref{transfg})
when $\bg$ is held fixed:
\be
\delta^{(Q)}_\epsilon \bg_{\mu\nu}=0\ ;
\qquad
\bg_{\mu\rho}\delta^{(Q)}_\epsilon (e^h)^\rho{}_\nu
=\Lie_\epsilon g_{\mu\nu}\ .
\ee
From the properties of the Lie derivative we have
\be
\Lie_\epsilon g_{\mu\nu}=\Lie_\epsilon \bg_{\mu\rho} (e^h)^\rho{}_\nu
+\bg_{\mu\rho}\Lie_\epsilon(e^h)^\rho{}_\nu
=(\bnabla_\rho\epsilon_\mu+\bnabla_\mu\epsilon_\rho) (e^h)^\rho{}_\nu
+g_{\mu\lambda}(e^{-h})^\lambda{}_\rho\Lie_\epsilon(e^h)^\rho{}_\nu
\ee
Then we find
\be
(e^{-h}\delta^{(Q)}_\epsilon e^h)^\mu{}_\nu=
(e^{-h}\Lie_\epsilon e^h)^\mu{}_\nu
+(e^{-h})^\mu{}_\rho(\bnabla^\rho\epsilon_\sigma+\bnabla_\sigma\epsilon^\rho)
(e^h)^\sigma{}_\nu
\ee
Expanding for small $h$ we find:
\be
\label{qtr}
\delta^{(Q)}_\epsilon h^\mu{}_\nu=
(\Lie_\epsilon\bar g)^\mu{}_\nu
+\Lie_\epsilon h^\mu{}_\nu
+[\Lie_\epsilon\bar g,h]^\mu{}_\nu
+O(\epsilon h^2)\ .
\ee
We note that the first two terms coincide with the quantum transformation when one uses
the linear background decomposition.
In the following we shall only be interested in the functional $\Gamma_k(h;\bar g)$
for $h=0$. It is therefore sufficient to consider only the first term in (\ref{qtr}).
Since $\delta^{(Q)}_\epsilon\bar g=0$ we can write
\be
\label{qtr2}
\delta^{(Q)}_\epsilon h_{\mu\nu}=
\bnabla_\mu\epsilon_\nu+\bnabla_\nu\epsilon_\mu+O(h)
\ee
Using the background $\bar g$ we can decompose the transformation
parameter $\epsilon^\mu$ in its longitudinal and transverse parts:
\be
\label{qpar}
\epsilon^\mu=\epsilon^{T\mu}+\bar\nabla^\mu \frac{1}{\sqrt{-\bar\nabla^2}} \psi\ ;
\qquad
\bar\nabla_\mu\epsilon^{T\mu}=0\ .
\ee
The inverse square root of the background Laplacian has been inserted
conventionally in the definition of $\psi$ so that it has the same
dimension as $\epsilon^\mu$.
We can then calculate the separate transformation properties
of the York-decomposed metric under longitudinal and transverse
infinitesimal diffeomorphisms.
We have
\be
\label{deltaxi}
\delta_{\epsilon^T}\xi^\mu=\epsilon^{T\mu}\ ;\qquad
\delta_\psi\omega=-\frac{1}{d}\sqrt{-\bar\nabla^2}\psi\ ;\qquad
\delta_\psi\sigma=\frac{2}{\sqrt{-\bar\nabla^2}}\psi\ ,
\ee
all other transformations being zero.
Note that $\sigma$ and $\omega$ are gauge-variant but the combination
$2\omega-\frac{1}{d}\bar\nabla^2\sigma$ is invariant.
In terms of the redefined variables (\ref{redef}) we have
\be
\label{deltaxiprime}
\delta_{\epsilon^T}\xi'_\mu=\sqrt{-\bar\nabla^2-\frac{\bar R}{d}}\,\epsilon^T_\mu\ ;\qquad
\delta_\psi\sigma'=2\sqrt{-\bar\nabla^2-\frac{\bar R}{d-1}}\,\psi\ .
\ee

First we pick the ``unimodular gauge'' $\det g=\det\bar g$.
\footnote{One often just sets $\det g=1$. This is incompatible with the
choice of a dimensionful metric, that we prefer.
Also note that on a compact manifold the gauge group does not allow one to make constant
rescalings of the metric, so that the overall scale of the metric remains a physical degree
of freedom. We ignore it in the following, since it does not affect the
running of the terms in our truncation.
}
In unimodular gravity this is imposed as an {\it a priori} condition
on the metric: by definition the path integral is then over
metrics with fixed determinant.
Here we start from the usual path integral over all metrics and
take $\det g=\det\bar g$ as a partial gauge condition.
This means that we have to take into account a ghost term.
To find the ghost operator we first observe that in the
exponential parametrization the unimodular gauge condition is 
\be
\label{omegagauge}
\omega=0\ .
\ee
From (\ref{deltaxi}) one then finds that the path integral must contain
a ghost determinant $\det(\sqrt{-\bar\nabla^2})=\sqrt{\det(-\bar\nabla^2)}$.
As usual, this can be rewritten as a path integral over a real anticommuting scalar ghost
\be
\label{omegaghost}
S_{g\omega}=\int d^dx\sqrt{\bar g}\, c(-\bar\nabla^2)c
\ee
(This gauge condition has been discussed previously in \cite{eichhorn}.
There, the ghost was a complex scalar.
This difference is due to the different definition of $\psi$ in (\ref{qpar}).)

The unimodular gauge condition completely breaks the invariance under longitudinal
infinitesimal diffeomorphisms, but leaves a residual gauge freedom that consists of 
the volume-preserving diffeomorphisms,
which are generated by the transverse vector $\epsilon^T$.
From (\ref{deltaxiprime}) we see that this residual freedom can be fixed by further choosing
\be
\label{xiprimegauge}
\xi'_\mu=0\ ,
\ee
which gives rise to a ghost determinant 
$\det\left(\sqrt{-\bar\nabla^2-\frac{\bar R}{d}}\right)
=\sqrt{\det\left(-\bar\nabla^2-\frac{\bar R}{d}\right)}$.
Again, this can be written as a path integral over an anticommuting real
transverse vector
\be
\label{xiprimeghost}
S_{g\xi}=\int d^dx\sqrt{\bar g}\, 
c_\mu\bar g^{\mu\nu}\left(-\bar\nabla^2-\frac{\bar R}{d}\right)c_\nu
\ee
Equations (\ref{omegagauge},\ref{xiprimegauge}) define the ``unimodular physical gauge'', 
which is the gauge condition that, unless otherwise stated, 
will be used in the rest of the paper.

Before proceeding it is instructive, however, to think for a moment of an alternative choice.
Since the combination $2\omega-\frac{1}{d}\nabla^2\sigma$ is gauge invariant,
one may alternatively also pick the gauge
\be
\label{sigmaprimegauge}
\sigma'=0\ .
\ee
From (\ref{deltaxiprime}) one then finds that the path integral must contain
a ghost determinant 
$\det\left(\sqrt{-\bar\nabla^2-\frac{\bar R}{d-1}}\right)
=\sqrt{\det\left(-\bar\nabla^2-\frac{\bar R}{d-1}\right)}$.
As usual, this can be rewritten as a path integral over a real anticommuting scalar ghost
\be
\label{sigmaprimeghost}
S_{g\sigma'}=\int d^dx\sqrt{\bar g}\, c\left(-\bar\nabla^2-\frac{\bar R}{d-1}\right)c
\ee
This choice may seem more natural but for our purposes it is less useful.
The reason is that if we set $\sigma'=0$, in the Hessian (\ref{decomposed})
there remains a kinetic term for $\omega$ which depends explicitly on $V$,
whereas if we set $\omega=0$ all kinetic operators are independent of $V$.
Since our purpose is precisely to avoid singularities due to the appearance
of $V$ in the kinetic operators, it is clear that for us here the second choice is preferable.

\subsection{Digression on Einstein-Hilbert gravity}

Since physical gauges are not very familiar, in this section we make a little 
digression to test our procedure in a setting that is better understood.
We consider the special case when $F$ and $V$ are constant,
in which case (\ref{newt}) can be used.
In this section we also drop the scalar field entirely.
The beta functions of the dimensionless couplings
$\tilde\Lambda=\Lambda/k^2$ and $\tilde G=Gk^{d-2}$, have the general form
\bea
\partial_t\tilde G&=&(d-2)\tilde G+B\tilde G^2\ ,
\\
\partial_t\tilde\Lambda &=&-2\tilde\Lambda+\frac{1}{2}A G+B\tilde G\tilde\Lambda\ ,
\eea
The coefficients $A$ and $B$ can be written as $A=A_1-\eta A_2$, $B=B_1-\eta B_2$, 
where $\eta$ is the anomalous dimension of $h_{\mu\nu}$.

We begin by considering only the case $\tilde\Lambda=0$
and we focus on the coefficient $B_1$.
We work in arbitrary dimension but we are especially interested
in the case $d=2+\epsilon$.
Let us first recall the situation with a standard gauge fixing term
\be
\label{alphabeta}
S_{GF}=\frac{1}{2\alpha}\int d^dx\sqrt{\bar g}\,
\bar g^{\mu\nu}\chi_\mu\chi_\nu\ ;
\qquad
\chi_\mu=\bar\nabla_\rho h^\rho{}_\mu-\frac{1+\beta}{d}\bar\nabla_\mu h\ ,
\ee
depending on two parameters $\alpha$ and $\beta$.
(The case $\alpha=1$, $\beta=d/2-1$ has been discussed recently in \cite{nink}.)
In the linear parametrization one finds, in the limit $d\to2$,
independently of $\alpha$,
\be
\label{b1lin}
B_1=-\frac{2\left(19-38\beta+13\beta^2\right)}{3(1-\beta)^2}
\ee
which for $\beta\to0$ gives $B=-38/3$,
a number that has been found many times in the literature 
\cite{tsao,brown,weinberg,kn,jack,reuter1}.
The same calculation with the exponential parametrization leads,
again independently of $\alpha$, to
\be
\label{b1exp}
B_1=-\frac{2\left(25-38\beta+19\beta^2\right)}{3(1-\beta)^2}
\ee
which for $\beta\to0$ reproduces the well-known
result of 2-dimensional quantum gravity $B_1=-50/3$ \cite{kkn,ddk,cdodo}.

What is one to make of this discrepancy?
We have computed the beta function of $G$ in generic dimension $d$,
where it is not universal, and then have taken the limit $d\to2$.
In this limit $G$ becomes dimensionless and it is generally the case
that the one-loop beta functions of dimensionless couplings are universal.
In fact the limits $d\to2$ do exhibit some degree of universality,
insofar as they can be shown to be independent of the choice of the cutoff function.
The choice of  parameterization does affect the limit, however.
The fact that the limit of the beta function does not generally agree with the
2-dimensional result is probably not too surprising, since our calculation
takes into account all the degrees of fredom of the metric, 
including the transverse traceless fluctuations, 
whereas in two dimensions those degrees of freedom do not exist, 
even at a kinematical level.
From this point of view it is perhaps surprising that 
one can reproduce the 2-dimensional result at all.
We observe that within the approach described here (one loop background-field
calculation in so called-single metric approximation)
the way to do so is to use the exponential parametrization
and the gauge $\beta=0$, whose special feature is that it does
not involve the conformal degree of freedom.
(For the limit it makes no difference whether one chooses
$\beta=0$ for all $d$, so that the conformal degree of freedom is not involved
in any dimension, or the de-Donder condition $\beta=d/2-1$,
where this is only true in $d=2$.)

Now let us see what happens in the physical gauges.
If we use a linear parametrization of the metric
then the gauge choice $\xi'=0$, $\sigma'=0$ gives $B_1=-38/3$.
In fact it can be seen that it gives the same result for $B_1$
as the standard gauge $\beta=0$, $\alpha=0$, in any dimension
and for any value of $\Lambda$.
This is because $\alpha=0$, $\beta=0$ means that we strongly
set to zero the quantity $\bar\nabla^\mu h^T_{\mu\nu}$.
On the other hand the gauge choice $\xi'=0$, $\omega=0$ gives $B_1=-26/3$.
We note that this coincides with the limit of (\ref{b1lin}) for $\beta\to\infty$.

If we use the exponential parametrization of the metric,
the gauge $\xi'=0$, $\omega=0$ gives $B_1=-38/3$ while
the gauge $\xi'=0$, $\sigma'=0$ gives $B_1=-50/3$.
\footnote{One has to be careful with the order of the limits.
Here we are always taking first $\tilde\Lambda\to0$ and then $d\to2$.
Taking the limits in the opposite order would give, in this case, $B_1=-52/3$.}
This calculation confirms the previous conclusion that in order to
reproduce the 2-dimensional result one has to use a gauge condition that
does not involve the conformal factor.
Since all these results depend critically on the form of the ghost terms,
we take their consistency as a confirmation of the correctness of the procedure.

We close this section by giving the complete beta functions of $\Lambda$ and $G$
in the exponential parametrization and in the physical gauges,
calculated with the optimized cutoff.
In the gauge $\xi'_\mu=0$, $\omega=0$ the coefficients of the beta functions are
(see \cite{cpr2} for other cutoff types)
\bea
A_1&=&\frac{16\pi(d-3)}{(4\pi)^{d/2}\Gamma[d/2]}
\\
A_2&=&-\frac{16\pi(d-1)}{(4\pi)^{d/2}(d+2)\Gamma[d/2]}
\\
B_1&=&- \frac{16\pi(d^5-4d^4-9d^3-48d^2+60d+24}{(4\pi)^{d/2}12 d^2(d-1)\Gamma[d/2]}
\\
B_2&=&\frac{16\pi(d^5-15d^3-58d^2+48}{(4\pi)^{d/2}12 d^2(d-1)(d+2)\Gamma[d/2]}
\eea
We will not attempt to calculate the anomalous dimension here.
If we restrict ourselves to the one-loop terms $A_1$ and $B_1$ there
is a Gaussian fixed point with critical exponents $2$ and $2-d$
and a nontrivial fixed point with critical exponents $d$ and $d-2$.
In $d=3$ the non-Gaussian fixed point is at $\tilde\Lambda=0$, $\tilde G=0.307$.
In $d=4$ it is at $\tilde\Lambda=0.133$, $\tilde G=3.35$.
Two features have to be stressed: the reality of the critical exponents
and the absence of singularities in the beta functions.
Both are due to the fact that the beta functions are independent of $\tilde\Lambda$.
The presence of singularities is usually attributed to deficiencies of the truncation
and there have been claims that the singularity should be replaced
by an infrared fixed point \cite{nagy,clpr}.
It is interesting to find at least one quantization procedure where this troublesome 
feature is absent even in the simple ``single-metric'' Einstein-Hilbert truncation.
\footnote{In \cite{cpr2} it was shown that the singularity is avoided also
with a ``type III'' or ``spectrally adjusted'' cutoff.
The price one had to pay there was a redefinition of what one means by high and
low momentum along the flow.
Here the same is achieved with a ``type I'' cutoff, where the distinction between
high and low momentum modes is conceptually cleaner, being based on the spectrum 
of a fixed second order differential operator.}

In the gauge $\xi'_\mu=0$, $\sigma'=0$ the beta functions depend on $\tilde\Lambda$
(via the kinetic operator of $\omega$).
They are much more complicated and we will only give here their form in $d=4$:
\footnote{In $d=3$ the non-Gaussian fixed point occurs at $\tilde\Lambda=0$, $\tilde G=0.257$
and has critical exponents $2.516$ and $1$.}
\bea
A_1&=&\frac{3-4\tilde\Lambda}{\pi(3-8\tilde\Lambda)}
\\
A_2&=&-\frac{-9+20\tilde\Lambda}{6\pi(3-8\tilde\Lambda)}
\\
B_1&=&\frac{-477+2448\tilde\Lambda-3392\tilde\Lambda^2}{24\pi(3-8\tilde\Lambda)^2}
\\
B_2&=&-\frac{171-888\tilde\Lambda+1280\tilde\Lambda^2}{27\pi(3-8\tilde\Lambda)^2}
\eea
We note that in this case there is a singularity at $\tilde\Lambda=3/8$.
The non-Gaussian fixed point occurs at $\tilde\Lambda=0.136$, $\tilde G=2.66$
and has complex critical exponents $2.44\pm0.593 i$.

\subsection{Flow equations}

The only effect of the gauge choice is to remove the four degrees of freedom
$\omega$ and $\xi$ from the theory and to add a real transverse vector ghost and a real scalar ghost.
The case of unimodular gravity is essentially identical, except that the scalar
ghost is not present.
(For recent related discussions of unimodular gravity see \cite{eichhorn,saltas}.)
The gauge fixed hessian is 
\bea
\label{gfhess0}
&&\int dx\sqrt{\bg}\Biggl[
F(\bphi)
\frac{1}{4}\tth_{\mu\nu}\left(-\bnabla^2+\frac{2\bR}{d(d-1)}\right)\tth^{\mu\nu}
-\frac{(d-1)(d-2)}{4d^2}F(\bphi)
\sigma'(-\bnabla^2)\sigma'
\nonumber\\
&&
-F'(\bphi)\frac{d-1}{d}\delta\phi
\sqrt{(-\bnabla^2)\left(-\bnabla^2-\frac{\bR}{d-1}\right)}\sigma'
+\frac{1}{2}\delta\phi\left(-\bnabla^2+V''(\bphi)-F''(\bphi)\bR\right)\delta\phi
\Biggr]
\eea
The field $\sigma'$ is invariant under volume-preserving diffeomorphisms,
so all three fields are physical.

As a final simplification we note that defining
\be
\sigma''=\sigma'+\frac{2d}{d-2}\frac{F'(\bar\phi)}{F(\bar\phi)}
\sqrt{\frac{-\nabla^2-\frac{\bar R}{d-1}}{-\nabla^2}}\delta\phi
\ee 
the gauge fixed hessian becomes diagonal:
\bea
\label{gfhess}
&&\int dx\sqrt{\bg}\Biggl[
F(\bphi)
\frac{1}{4}\tth_{\mu\nu}\left(-\bnabla^2+\frac{2\bR}{d(d-1)}\right)\tth^{\mu\nu}
-\frac{(d-1)(d-2)}{4d^2}F(\bphi)
\sigma''(-\bnabla^2)\sigma''
\nonumber\\
&&
+\frac{1}{2}\delta\phi\left(-\bnabla^2+V''(\bphi)-F''(\bphi)\bR +2\frac{d-1}{d-2}\frac{F'(\bphi)^2}{F(\bphi)}\left(-\bnabla^2-\frac{\bR}{d-1}\right)\right)\delta\phi
\Biggr]\,.
\eea

From here on we proceed as in \cite{narain} and for notational simplicity we shall 
remove the bars from the background fields.
We choose the cutoff in such a way that the modified inverse propagator
is identical to (\ref{gfhess}) except for the replacement of
$-\bar\nabla^2$ by $P_k(-\bar\nabla^2)=-\bar\nabla^2+R_k(-\bar\nabla^2)$.
We note that applying this procedure directly to Eq.~\ref{gfhess0},
as in \cite{narain}, would amount to a slightly different definition of the cutoff.
Both procedures seem legitimate, and our choice is dictated purely by later convenience.
\footnote{Neglecting $\dot F$ on the r.h.s. of the flow equation, the fixed point equations derived
from the two procedures turn out to be the same.}

The flow equations for $V$ and $F$ can be extracted from
(\ref{erge}) using the formula
\be
\label{HKasymp}
{\rm Tr}W(-\bnabla^2)=\frac{1}{(4\pi)^{d/2}}
\left[Q_{\frac{d}{2}}(W)B_0(-\bnabla^2)
+Q_{\frac{d}{2}-1}(W)B_2(-\bnabla^2)+\ldots\right]\ ,
\ee
where the coefficients $Q_n(W)$ are given, for $n>0$, by
$Q_{n}(W)=\frac{1}{\Gamma(n)}\int_{0}^{\infty}dz\, z^{n-1}W(z)$.
Since we keep at most terms linear in $R$, we need only the
first two terms of the expansion of the heat kernel of $-\bnabla^2$.
We refer to the appendices in \cite{cpr2} for a pedagogical discussion.
With standard procedure, neglecting in our LPA truncation the anomalous dimension 
of the scalar field, one arrives at the flow equations for the dimensionless functions 
of the dimensionless field $\varphi=k^{\frac{2-d}{2}}\phi$:
$f(\varphi)=k^{2-d}F(\phi)$ and $v(\varphi)=k^{-d}V(\phi)$.

We shall consider two approximation schemes.  
As a first case we neglect derivatives of $F_k$ with respect to $k$ 
in the r.h.s. of the flow equation.
The analysis of the scaling solutions of the resulting equations, and their
eigenperturbations, for $d=3$ and $d=4$ is given in Sections IV and V.
With the insight obtained in this way, in section VI we shall
consider the full equation where the terms proportional to $\dot F$ are not neglected.
This means that we replace $\partial_t f\to 0$
in the r.h.s. of the fixed point equation (but not of the flow equation,
hence also not in the analysis of eigenperturbations). 
The discussion here will be short.
In the rest of the paper we consider only the cases $d=3$ and $d=4$.
In appendix C we shall give the form of the flow equations for general dimension. 

If we neglect $\dot F$, in the r.h.s. the flow equations in $d=3$ read
\bea
\label{vdot3}
\dot v&=&
-3\,v+\frac{1}{2}\varphi\,v'+
\frac{f+4f'^2}{6\pi^2\left(f(1+v'')+4f'^2\right)}
\\
\label{fdot3}
\dot f&=&
-f+\frac{1}{2}\varphi\,f'+\frac{37}{72\pi ^2}
+f\frac{(f+4f'^2)(2-4 f''+5v'')+3 f v''^2}{24\pi^2(f(1+v'')+4f'^2)^2}
\eea
whereas in $d=4$
\bea
\label{vdot4}
\dot v&=&
-4\,v+\varphi\,v'+\frac{1}{16\pi^2}+
\frac{f+3f'^2}{32\pi^2\left(3f'^2+f(1+v'')\right)}
\\
\label{fdot4}
\dot f&=&
-2f+\varphi\,f'+\frac{37}{384\pi ^2}
+f\frac{(f+3f'^2)(1-3 f''+3v'')+2 f v''^2}{96\pi^2(3f'^2+f(1+v''))^2}
\eea

It is interesting to note that if we assume $f'=0$, 
and discard the constant terms in the r.h.s, the equation for $v$
reduces to the flow equation (\ref{vdotscalar}) for the potential
in pure scalar theory, derived with the same cutoff.
In general, shifting the r.h.s. of (\ref{vdotscalar}), (\ref{vdot3}) or (\ref{vdot4}) by a constant
results only in a constant shift of the potential of the solution, 
and is therefore immaterial in flat space.
The constants in the r.h.s. of (\ref{fdot3}) and (\ref{fdot4}) 
cannot be discarded in the same way because
these equations contains $f$ also in the nonlinear part.

\section{Scaling solutions in $d=3$}

\subsection{Analytic solutions}

Let us look for solutions of the system (\ref{vdot3},\ref{fdot3}).
Assuming that $v$ and $f$ are constant, there is a unique solution
\be
\label{gmfp}
v_*=\frac{1}{18\pi^2}\approx0.005629\ ;
\qquad
f_*=\frac{43}{72\pi^2}\approx 0.06051\ .
\ee
In \cite{narain} this was called the ``Gaussian Matter Fixed Point'',
since at this fixed point the EAA becomes quadratic in $\phi$.
\footnote{As pointed out in \cite{eichhorn2}, graviton loops will generally
induce derivative interactions in the matter sector,
so the existence of this fixed point is limited to
a truncation where such interactions are absent.}
Here we shall call it ``FP1''.
As in (\ref{newt}) we change variables to
\be
v=\frac{2\tilde\Lambda}{16\pi\tilde G}\ ;
\qquad
f=\frac{1}{16\pi\tilde G}\ .
\ee
whose beta functions are
\bea
\label{vdoteh}
\beta_{\tilde G}&=&
\tilde G-\frac{86}{9\pi}\tilde G^2
\\
\label{fdoteh}
\beta_{\tilde\Lambda}&=&
-2\tilde\Lambda
-\frac{86}{9\pi}\tilde G\tilde\Lambda
+\frac{4}{3\pi}\tilde G\ .
\eea
In these variables one sees the Gaussian fixed point at  $\tilde G=\tilde\Lambda=0$
(which cannot be seen in the other variables) and FP1 at
\be
\tilde\Lambda_*=\frac{2}{43}\approx 0.04651\ ;
\qquad
\tilde G_*=\frac{9\pi}{86}\approx 0.3288\ .
\ee
The difference with the values given in section III.C is due to the
presence of the scalar field in the loops.
These values are somewhat larger than the ones that had been found
for pure gravity in three dimensions either in the Einstein-Hilbert truncation,
or with a Chern-Simons term \cite{ps,ppps}, or with higher derivative terms \cite{ohta}.
\bigskip

For this fixed point it is possible to perform an analytical study of the eigenperturbations
in the infinite dimensional functional space spanned by $v$ and $f$.
The general linear equations for the eigenperturbations are of the form
$(\hat O-\lambda)\delta w=0$,
where $\delta w^T=(\delta v, \delta f)$ and $\hat{O}$ is the corresponding differential operator,
which is constructed by substituting into the fixed point equations
\be
v(\varphi)=v_*(\varphi)+\epsilon\delta v(\varphi) e^{\lambda t} \quad , \quad 
f(\varphi)=f_*(\varphi)+\epsilon\delta f(\varphi) e^{\lambda t}\ ,
\ee
and expanding to first order $\epsilon$. 
Linearization around FP1 leads to the equations
\bea
&{}&-(\lambda +3)\delta v +\frac{\varphi}{2} \delta v'-\frac{\delta v''}{6 \pi ^2}=0 \nonumber\\
&{}&-(\lambda +1)\delta \!f+\frac{\varphi}{2}\delta \!f'-\frac{\delta \!f''}{6 \pi ^2} +\frac{\delta
   v''}{24 \pi ^2}=0
\label{eigend3eq1}
\eea
which can be studied both analytically and numerically.
Imposing that the eigenfunctions are even functions of $\varphi$ leads to a quantization condition 
for the eigenvalues: $\lambda=-\theta=-3,-2,-1,0, ...$
(with $\theta=-\lambda$ the critical exponent). The relevant perturbations
(those with $\theta>0$) are
\bea
&{}&\theta_1=3 , \quad \quad w_1^t=(\delta v,\delta\! f)_1 = (1,0) 
\nonumber\\
&{}&\theta_2=2 , \quad\quad w_2^t=(\delta v,\delta\! f)_2 = 
\left(-\frac{1}{3\pi^2}+\varphi^2,1\right) 
\nonumber\\
&{}&\theta_3=1 , \quad\quad w_3^t=(\delta v,\delta\! f)_3 = (0,1),
\nonumber\\
&{}&\hspace{1.2cm}  
\quad\quad w_4^t=(\delta v,\delta\! f)_4 = \left(\frac{1}{3\pi^4}-\frac{2}{\pi^2}\varphi^2+\phi^4,-\frac{1}{2\pi^2}\varphi^2\right)
\nonumber\\
&{}&\theta_4=0 , \quad\quad w_5^t=(\delta v,\delta\! f)_5 = 
\left(0,-\frac{1}{3\pi^2}+\varphi^2\right) , \nonumber\\
&{}&\hspace{1.2cm} \quad\quad w_6^t=(\delta v,\delta\! f)_6 = \left(-\frac{5}{9\pi^6}+\frac{5}{\pi^4}\varphi^2-\frac{5}{\pi^2}\varphi^4+\varphi^6,
\frac{5}{4\pi^4}\varphi^2-\frac{5}{4\pi^2}\varphi^4\right) 
\eea

Note that the eigenspaces of the eigenvalue $-1$ and $0$ are two-dimensional.
The asymptotic behavior of $w_6$ as well as of all the eigenperturbations with eigenvalues $\lambda>0$ is power-like.
We have computed also some eigenperturbations associated to the irrelevant directions but we do not find it necessary to give their expressions here.
Moreover it can be shown that all the eigenperturbations found are not redundant.
A check with a numerical eigenperturbation analysis has been done using a spectral method based on Chebyshev polynomials
and leads to the same result. We shall use the numerical approach later on when the analytical approach is not possible.

There is then another analytic solution with constant $v$ but nonconstant $f$:
\be
\label{other}
v_*=\frac{1}{18\pi^2}\approx 0.005629\ ;\qquad 
f_*=\frac{37}{72\pi^2}+\frac{1}{4}\varphi^2\approx0.0520678+0.25\varphi^2\ ,
\ee
which corresponds to
$$
\tilde G_*=\frac{9\pi}{74}\approx 0.3821\ ;\qquad 
\tilde\Lambda_*=\frac{2}{37}\approx 0.05405\ ;\qquad
\xi=\frac{1}{4}
$$
(where we follow the standard terminology of calling $\xi$ 
the coefficient of the nonminimal coupling $\phi^2 R$).
We shall call this fixed point FP2.
We emphasize that although this solution only has three nonvanishing couplings,
it is a solution of the full functional equations, not of a truncated subset.
\bigskip

The linearized flow equations around FP2 are:
\bea
&{}&- (\lambda +3)\delta v+\frac{\varphi}{2} \delta v'-\frac{\left(18 \pi ^2 \varphi ^2+37\right)}{6 \pi ^2 \left(90 \pi ^2 \varphi ^2+37\right)} \delta v''=0 
\nonumber\\
&{}&- (\lambda +1)\delta \!f+\frac{\varphi}{2}  \delta\! f'-\frac{\left(18 \pi ^2 \varphi ^2+37\right)}{24 \pi ^2 \left(90 \pi ^2 \varphi ^2+37\right)} \left(4\delta \!f''-5\delta v''\right)=0
\label{eigeneq1}
\eea

\begin{figure}
\includegraphics[width=6cm]{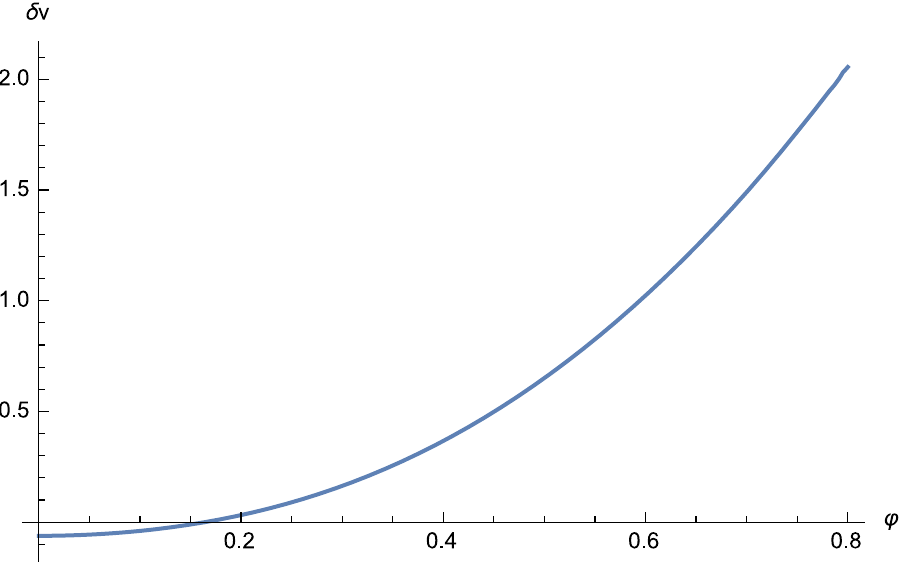}
\qquad
\includegraphics[width=6cm]{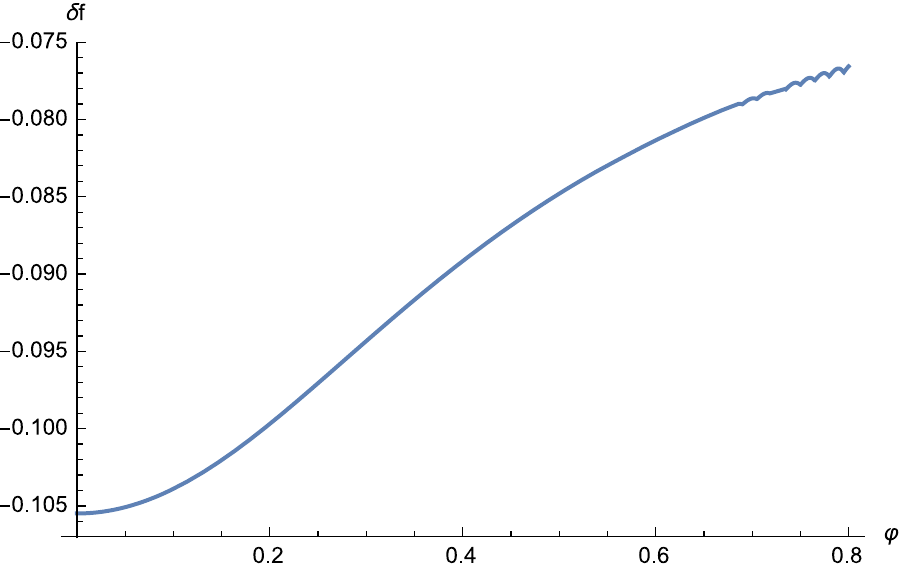}
\\
\bigskip
\includegraphics[width=6cm]{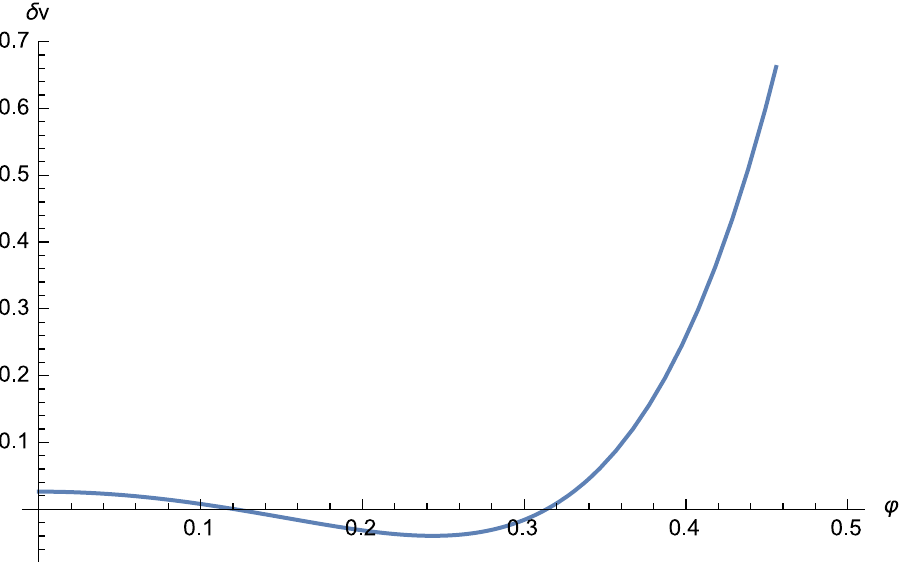}
\qquad
\includegraphics[width=6cm]{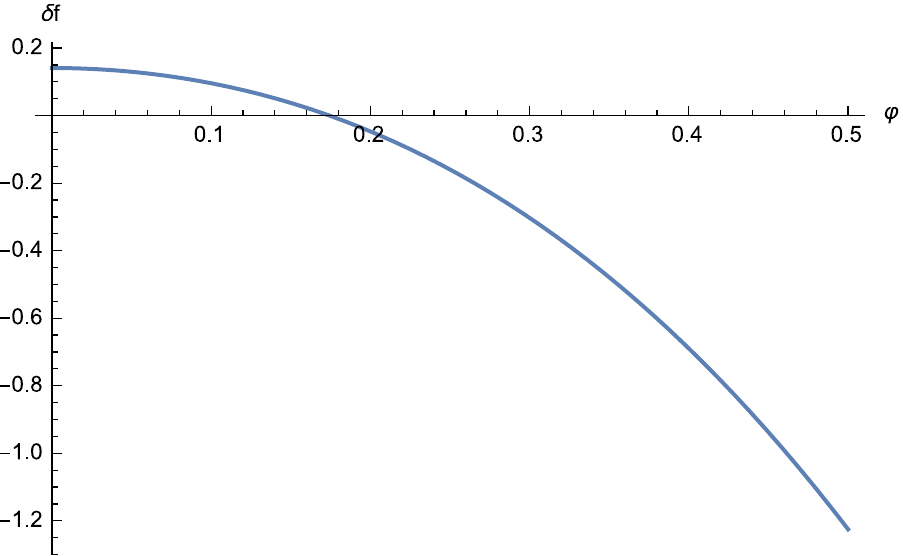}
\caption{Top: the eigenperturbations $\delta v_2$ and $\delta \!f_2$ for $\theta=1.815$.
Bottom: the eigenperturbations $\delta v_4$ and $\delta \!f_4$ for  $\theta=0.516$.}
\label{eigen2plot}
\end{figure}
\medskip
We proceed here with numerical methods which we have tested with the previous case. The numerical approximation for the eigenfunctions is good for small values of $\varphi$ but this 
is enough for our purposes.
We find four relevant direction. The critical exponents and corresponding approximated eigenpertubations (rewritten as simple polynomials in a small $\varphi$ range around the origin)
\bea
&{}&\theta_1=3 , \quad \quad  \ \ w_1^t=(\delta v,\delta\! f)_1 = (1,0) \nonumber\\
&{}&\theta_2=1.815 , \quad  \! w_2^t=(\delta v,\delta\! f)_2  \nonumber\\
&{}&\theta_3=1 , \quad \quad \ \ w_3^t=(\delta v,\delta\! f)_3 = (0,1)\nonumber\\
&{}&\theta_4=0.516 , \quad w_4^t=(\delta v,\delta\! f)_4
\eea
We show in Fig.~\ref{eigen2plot} in different plots the components of $w_2$ and $w_4$.

Finally there is a fixed point that we shall call FP3:
\be
\label{other}
v_*=\frac{1}{18\pi^2}\approx 0.005629\ ;\qquad 
f_*=-\frac{43}{568}\varphi^2\approx -0.0757\varphi^2\ ,
\ee
where Newton's coupling, defined in terms of $f[0]$, is formally infinite
and $f$ is otherwise always negative.
We shall not give the results for the fluctuations around this fixed point here.

\subsection{Search for gravitationally dressed Wilson-Fisher FP}

Finally we look for a solution with nontrivial $v$ and $f$.
We begin by considering finite polynomial truncations
\be
\label{poly}
v(\varphi)=\sum_{i=0}^{n_v}\lambda_{2i}\varphi^{2i}\ ;\qquad
f(\varphi)=\sum_{i=0}^{n_f}\xi_{2i}\varphi^{2i}\ .
\ee
We have considered polynomials in $\varphi^2$ of order up to 9 for $v$ and 8 for $f$.
At each order one finds a fixed point whose features are close to those of
a fixed point in the lower truncation, so we assume that these are the trace
of a genuine fixed point.
For the highest truncation the fixed point solution is
\bea
v(\varphi)&=&0.00679-0.0856\varphi^2+0.568\varphi^4+1.353 \varphi^6+2.903 \varphi^8+3.390 \varphi^{10}
\nonumber
\\
&& -7.737\varphi^{12}-55.00\varphi^{14}-122.8\varphi^{16}+30.45\varphi^{18}
\\
f(\varphi)&=& 
0.0625-0.0578\varphi^2-0.0608\varphi^4+0.00270\varphi^6+0.318 \varphi^8 + 0.675\varphi^{10}
\nonumber
\\
&&
-2.411\varphi^{12}-20.53\varphi^{14}-52.61\varphi^{16}\ .
\eea
This already shows the striking difference between the equations obtained here and
the ones of \cite{narain}, that only admitted potentials unbounded from below.

The fixed potential has $v(0)=0.00679$ and a nontrivial minimum at $\varphi\approx 0.245$
equal to $v\approx0.040$.
Although the fixed point is encouragingly stable as one increases the order of
the polynomials, the radius of convergence of this series expansion does
not extend much further than the nontrivial minimum.
One can get a slightly improved series by expanding around the minimum
instead of zero.
In this case we obtain
\bea
\label{vmin}
v(\varphi)&=&0.0040 + 0.8877 (\varphi^2-\kappa)^2 
+ 2.177 (\varphi^2-\kappa)^3 
+3.451 (\varphi^2-\kappa)^4  
\nonumber\\
&&
-1.653 (\varphi^2-\kappa)^5
-20.03 (\varphi^2-\kappa)^6 
+12.10 (\varphi^2-\kappa)^7 
+283.9 (\varphi^2-\kappa)^8 
\\
&&
-53,68 (\varphi^2-\kappa)^9 
-5592 (\varphi^2-\kappa)^{10} 
-6163 (\varphi^2-\kappa)^{11} 
+12691 (\varphi^2-\kappa)^{12}\ ;
\nonumber\\
\label{fmin}
f(\varphi)&=& 0.0588
-0.06488(\varphi^2-\kappa)
-0.05302(\varphi^2-\kappa)^2
+0.08890 (\varphi^2-\kappa)^3
+0.3073(\varphi^2-\kappa)^4 
\nonumber\\
&&
-1.073(\varphi^2-\kappa)^5
-5.403(\varphi^2-\kappa)^6
+17.25(\varphi^2-\kappa)^7
+126.0(\varphi^2-\kappa)^8 
\\
&&
-318.6(\varphi^2-\kappa)^9
-4270(\varphi^2-\kappa)^{10}
+1088(\varphi^2-\kappa)^{11}
\nonumber\ ,
\eea
where $\kappa=\varphi_{min}^2=0.0600$, in agreement with the expansion around the origin.
This expansion reaches slightly larger values of the field than the expansion around zero.

It is instructive to compare these approximate solutions
to the solution of the equation (\ref{vdotscalar}),
which is an approximation of the Wilson-Fisher fixed point
in the Local Potential Approximation. 
When compared, the two curves are practically indistinguishable,
up to a little beyond the minimum of the potential.
This can be understood by noting that
if $f'=0$ the equation (\ref{vdot3}) reduces to (\ref{vdotscalar}). 
Thus the difference between the two solutions is entirely due to the fact that
$f$ is not constant.

One can calculate the critical exponents in this polynomial approximations
by taking the eigenvalues of the $(n_v+n_f)\times(n_v+n_f)$ matrix of derivatives 
of the beta functions with respect to the couplings.
The best estimates give the critical exponents:
$3$, $1.562$, $0.997$, $-0.588\pm0.108 i$\ldots
The first eigenvector points precisely in the direction of $\lambda_0$,
the others are admixtures of all couplings.

\begin{figure}
\includegraphics[width=8cm]{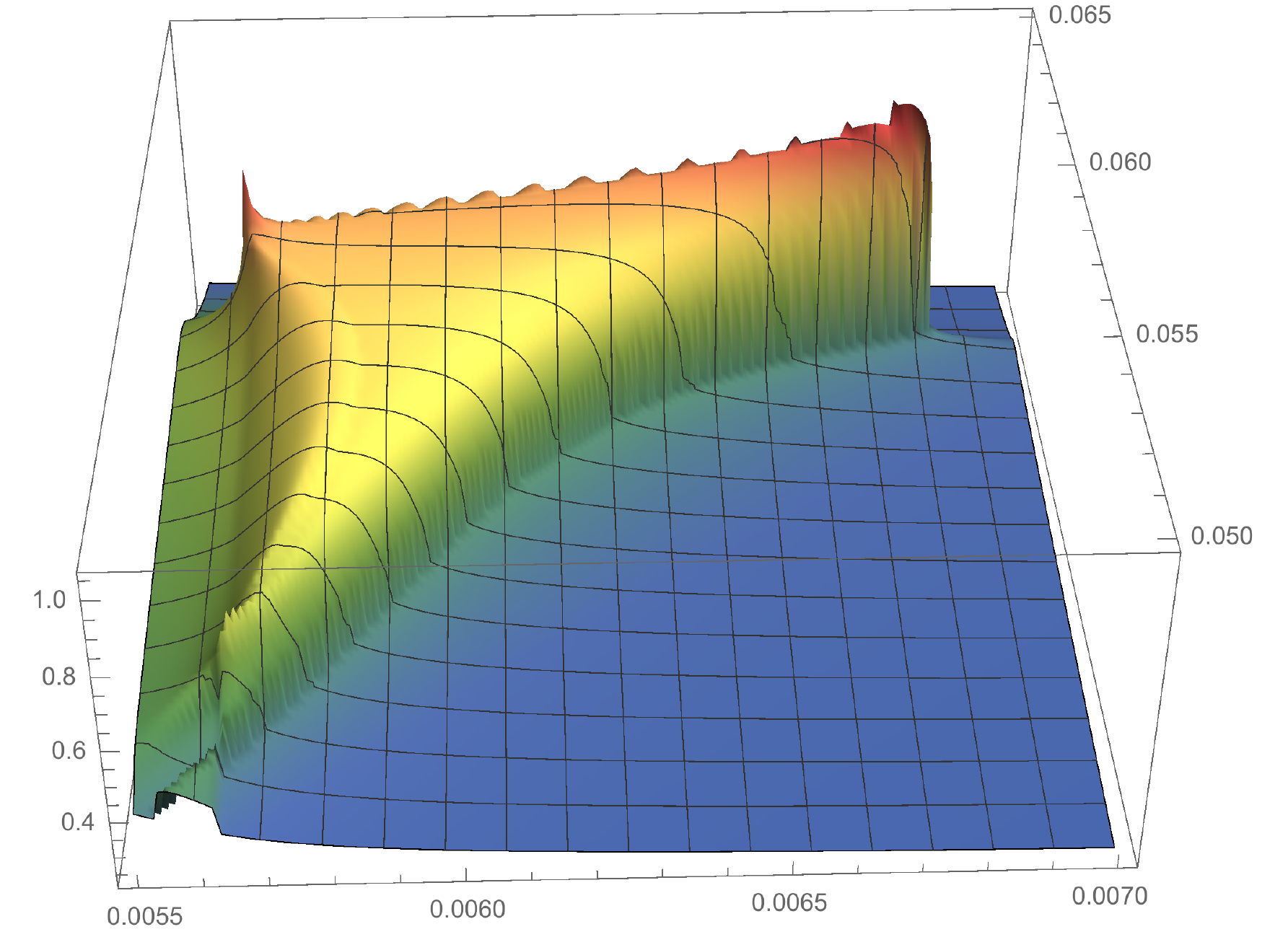}
\caption{Plot of the maximum value of $\varphi$ reached by the numerical integrator in the most interesting region  of the initial conditions 
$0.0055<v(0)<0.0070$ and $0.050<f(0)<0.065$.
The jagged appearance of the top ridge is a numerical artifact.}
\label{spike_d3_v0_f0}
\end{figure}

\begin{figure}
{\resizebox{0.95\columnwidth}{!}
{\includegraphics{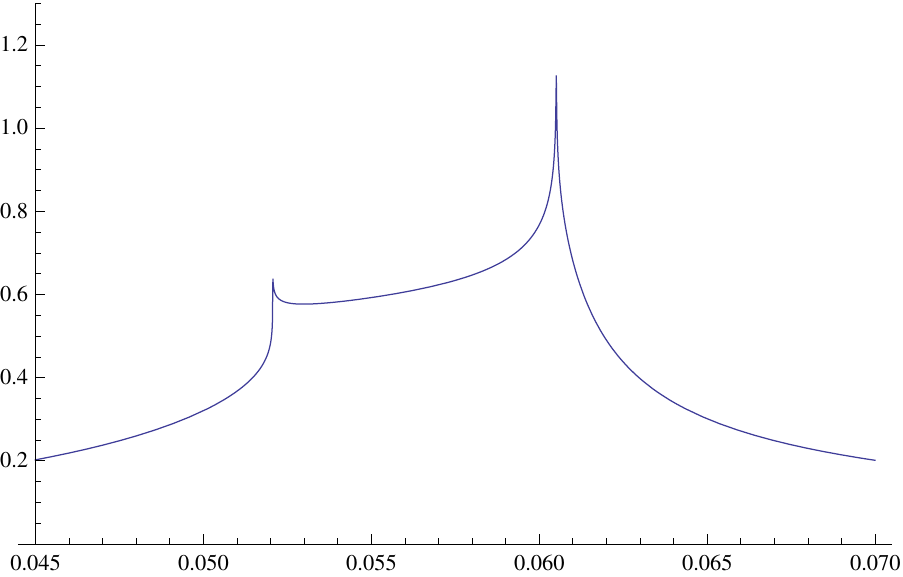}\qquad
\includegraphics{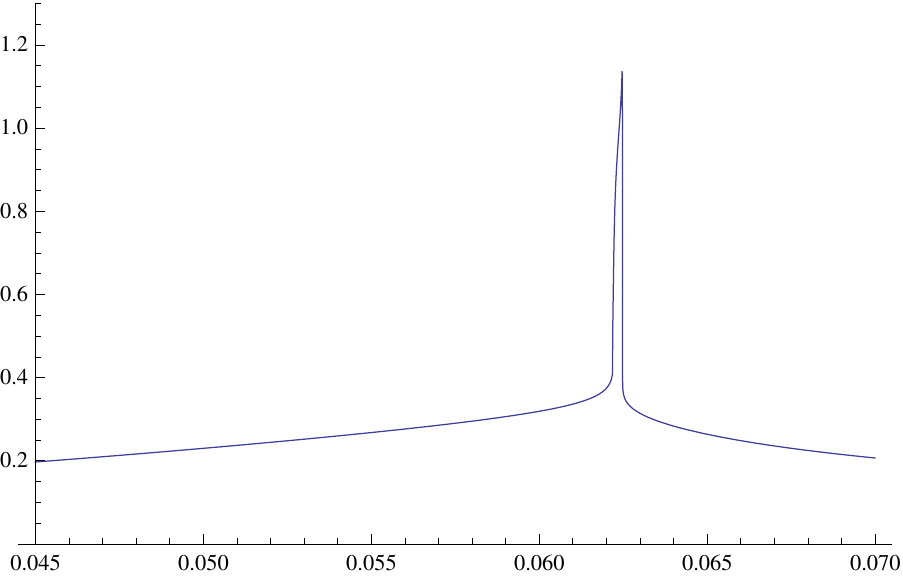}
}}
\caption{Left panel: Spike plot as function of $f(0)$ for $v(0)=1/18\pi^2\approx 0.0056$.
The peak on the right is FP1 (\ref{gmfp}), 
the one on the left is FP2 (\ref{other}).
Right panel: Spike plot as function of $f(0)$ for $v(0)=0.0068$.
There is no clear spike for intermediate values of $v(0)$.
}
\label{spike-cuts}
\end{figure}

These finite truncations can be complemented by numerical studies along the
lines of \cite{morris2}.
One tries to solve numerically the fixed point equations
$\dot v=0$, $\dot f=0$ with the initial conditions $v'(0)=0$ and $f'(0)=0$.
One can then plot how far the numerical routines can go,
as a function of the initial conditions $v(0)$ and $f(0)$.
Fixed points then typically appear as spikes in this graph.
We have charted an area around the origin in the $v(0)$-$f(0)$ plane.
There is a large region for $v(0)<0$ and $f(0)>0$ where the solution easily extends
up to arbitrarily large $\varphi$.
These solutions all have potentials that are unbounded from below
(they behave asymptotically as in equation (\ref{altas}) below).
For $f(0)<0$ there are areas where the behavior look quite chaotic.
We cannot say much about the system for such initial conditions.
The most interesting area is a mountainous triangle in the quadrant $v(0)>0$ and $f(0)>0$,
enlarged in Fig.~\ref{spike_d3_v0_f0}.
It is relatively smooth but the ridges become quite sharp near its vertices and
there are distinct peaks at the end of each ridge.
Two of these can be seen clearly if we cut the graph along the line $v(0)=1/(18\pi^2)$, 
which is common to the two fixed points (\ref{gmfp},\ref{other}).
Then the two fixed points appear as very clear spikes at
initial conditions that agree numerically with the values
of $\lambda_0$ and $\xi_0$ of the polynomial solutions,
as well as the ``exact'' values.
\footnote{We note in passing that the fixed point FP3, which also lies on the same line,
cannot be seen by this technique because it corresponds to the initial condition $f(0)=0$
where the equations in normal form have a singularity.
For the same reason this solution also does not show up among the solutions
of the polynomial expansion around $\varphi=0$.}
The nontrivial fixed point can be seen by cutting along
the line $v(0)=0.0068$, as seen in Fig.~\ref{spike-cuts}.
One should be careful in interpreting such plots.
They probably reveal as much about the workings of the numerical
integration algorithm than about the equations themselves.
Nevertheless, the coincidence of the results from polynomial truncations
and numerical integration is good evidence for the existence of fixed points.
Two of these can also be seen analytically, so the main issue is whether
the third one can be continued to large $\varphi$.

In order to understand the behavior of the solution for large field, let us
write the fixed point equations 
$\dot v=0$ and $\dot f=0$ in normal form:
\bea
\label{vdotn3}
v''&=&
\frac{(f+4f'^2)\left(1-3\pi^2(6v-\varphi v')\right)}{3 \pi ^2 f(6v-\varphi v')}
\\
\label{fdotn3}
f''&=&
\frac{1}{108\pi^4 f(6v-\varphi v')^2}
\big(
-216\pi^4f'^2\left(6v-\varphi v'\right)^2-9\pi^2(f+4f'^2)(6v-\varphi v')
\nonumber\\
&&
+\left(36\pi^2\varphi f'-72\pi^2f+46\right)(f+4f'^2)\big)
\eea
If we assume that for large $\varphi$, $1$, $f$, $\varphi f'$ and $f'^2$ are all
negligible with respect to $6v-\varphi v'$, then the solution would behave
asymptotically like 
\be
\label{altas}
v\sim-\frac{1}{2}\varphi^2+C_1\varphi+C_2+2C_3(3\varphi-C_4)^{1/3}\ ;\qquad
f\sim C_3(3 \varphi-C_4)^{1/3}\,.
\ee
They are analogous to the solutions of the pure scalar flow equation $\dot v=0$
with $\dot v$ given by (\ref{vdotscalar}), with initial condition $v(0)<0$,
or equivalently $v''(0)<-1$.
Such solutions exist for all $\varphi$ but are unbounded from below and are
therefore unphysical.

Another possible asymptotic behavior, which is expected for dimensional reasons,
is $v(\varphi)=A\varphi^6+\ldots$ and $f=B\varphi^2+\ldots$, for some constants $A$ and $B$.
This is the behavior that leads to the Wilson-Fisher fixed point in the pure scalar case.
In order to better approximate the solution for large field
one can use an expansion in $1/\varphi$.
The subleading terms can be calculated iteratively from the equations (\ref{vdot3},\ref{fdot3}).
One finds
\footnote{In actual calculations we have pushed this expansion to order $1/\varphi^{40}$.}
\bea
\label{asyv}
v(\varphi)&\!\!=&\!\!A \varphi^6+\frac{1}{9\pi^2}
+\frac{16 B+1}{900\pi^2 A\varphi^4}
-\frac{19}{2430\pi^4 A\varphi^6}
-\frac{6912\pi^4 B^3\!+\!864\pi^4 B^2\!+\!27\pi^4 B\!-\!3610 A}{1020600\pi^6 A^2 B\varphi^8}
+\ldots
\\
\label{asyf}
f(\varphi)&\!\!=&\!\! B\varphi^2+\frac{23}{36\pi^2}
-\frac{16 B+1}{2160\pi^2 A\varphi^4}
+\frac{23}{6480\pi^4 A\varphi^6}
-\frac{10368\pi^4 B^3\!+\!216\pi^4 B^2\!-\!27\pi^4 B\!+\!5290A}{2916000\pi^6 A^2 B \varphi^8}
+\ldots
\eea

There are corrections to this behavior which can be  obtained by studying the linearized differential equations around such an asymptotic solution.
One can see that there are four linear independent solutions to the linearized equations, but two of them just renormalize the constants $A$ and $B$ of the previous expansion,
while the other two are truly new corrections with an essential singularity in the variable $1/\varphi$. 
We find that both $\delta v$ and  $\delta \! f$ are proportional to
\be
 \exp \left(-\frac{\varphi ^{10}}{20 c}+\frac{23 b^2 \varphi ^8}{1536 c^2}+\ldots\right) \cosh \left(\frac{\sqrt{b} }{14 \sqrt{2} c} \varphi ^7+\ldots \right)
 \label{linasy}
\ee
where $b=\frac{4}{45 \pi ^2 A}$ and $c=-\frac{1+16 B}{5400 \pi ^2 A^2}$
and differ starting from the power-like term, i.e. in the $\log$ term inside the exponential.

Equations(\ref{vdot3},\ref{fdot3}) do not have fixed singularities,
but they can have movable singularities.
At such points $\varphi_s$, both $v''$ and $f''$ become singular:
\bea
\label{vsing}
v(\varphi)&=&(\varphi -\varphi_s)^{3/2} \left[ A_0+O(\varphi -\varphi_s) \right]+v_0+v_1 (\varphi -\varphi_s)+O\left( \varphi-\varphi_s\right)^2 \nonumber \\
\label{fsing}
f(\varphi)&=&(\varphi -\varphi_s)^{3/2} \left[ B_0+O(\varphi -\varphi_s) \right]+f_0+f_1 (\varphi -\varphi_s)+O\left( \varphi-\varphi_s\right)^2
\eea
with
\bea
A_0=-8 B_0 \,, \ \ B_0=\frac{\sqrt{9 \pi ^2 f_0 \left(184-9 \pi ^2 \varphi_s^2\right)\!-\!1296 \pi ^4 f_0^2\!-\!529}}{54 \sqrt{6} \pi ^3 \sqrt{f_0}\,
   \varphi_s^{3/2}}\,, \ \ v_1=6 \frac{v_0}{\varphi_s}\, \ \ f_1=\frac{36f_0\!-\!\frac{23}{\pi^2}}{18\varphi_s}
   \nonumber
\eea
We have not investigated further the equations for other possible singular behaviors.

If the solution does not end at such a singularity,
it should be possible to match the expansion for small field to the expansion for large field. 
In the pure scalar case this method leads to a very good overlap of the approximate solutions
and to good estimates of critical exponents \cite{morris3,LZ}.
In the present case this does not happen: the radii of convergence of the expansions do not overlap.
We have tried to extend the polynomial solutions towards the right by selecting
suitable Pad\'e approximant.
It is then possible to adjust the free parameters $A$ and $B$ of the large field 
expansion to match quite well the small field Pad\'e approximants.
We find the best match for $A\approx3.41$, $B\approx-0.144$.
Unfortunately this is misleading: neither the Pad\'e approximant not the asymptotic expansion
are good approximations of the solution near that point.
The reason is that the Pad\'e approximant of $f$ has a zero at a point $\varphi_0\approx 0.9$, 
and the equation is singular for $f=0$,
as best seen from the normal forms (\ref{vdotn3},\ref{fdotn3}).

\begin{figure}
\includegraphics[width=6cm]{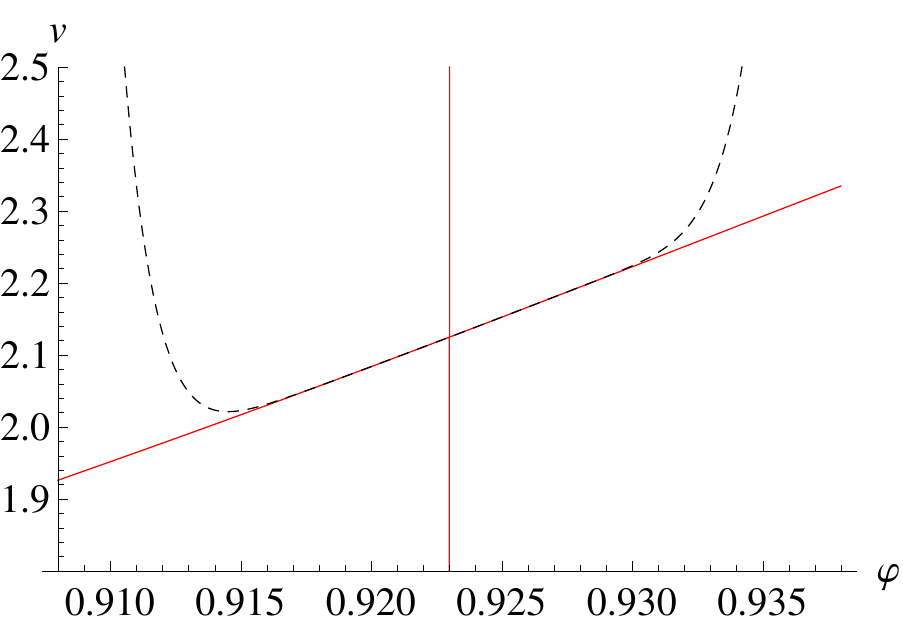}
\qquad
\includegraphics[width=6cm]{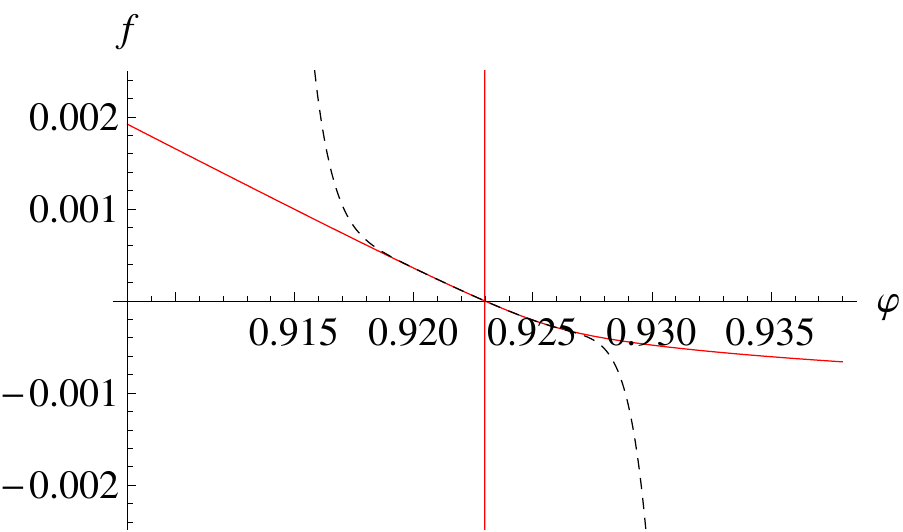}
\\
\bigskip
\includegraphics[width=6cm]{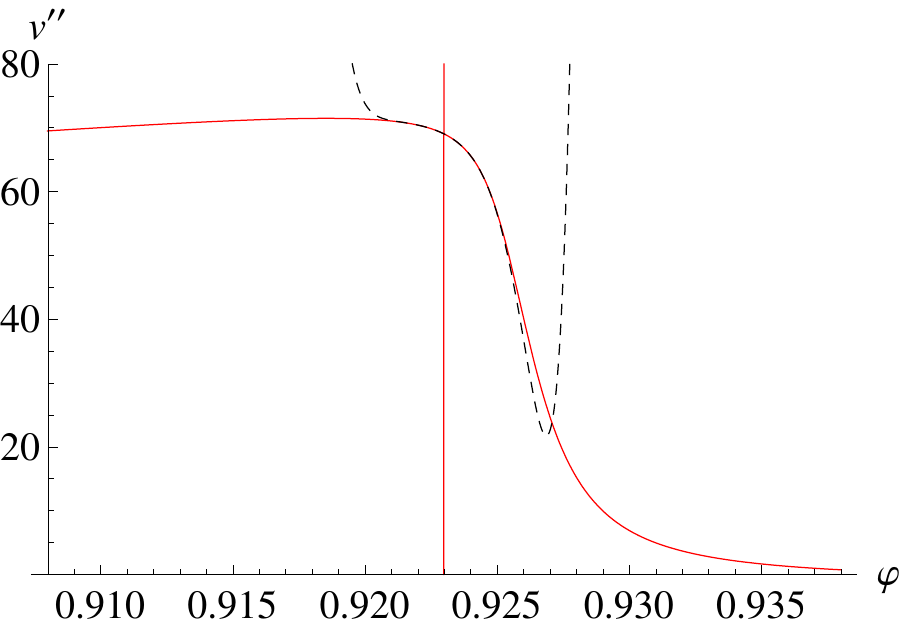}
\qquad
\includegraphics[width=6cm]{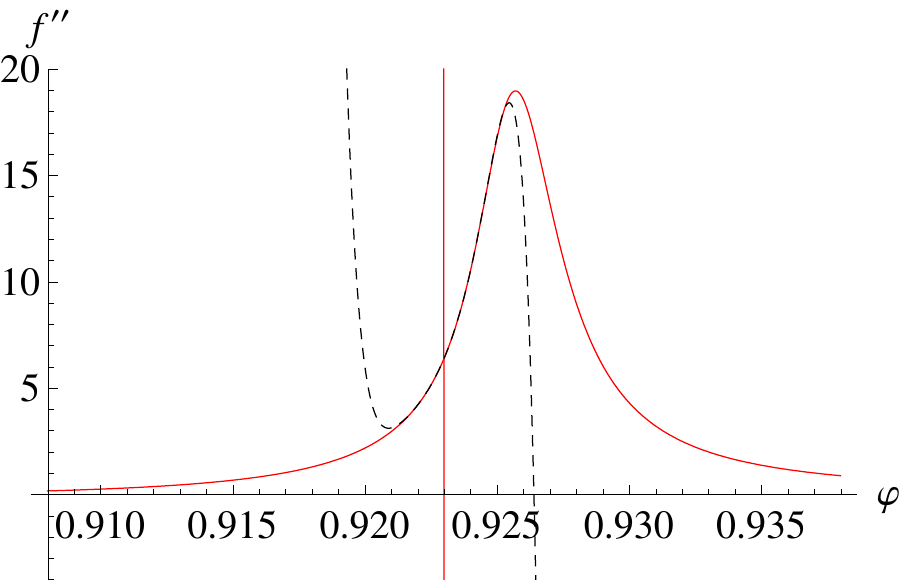}
\caption{Enlargement of the solution near $\varphi_0$. 
Top: $v$ and $f$.
Bottom: their second derivatives.
Red continuous curve: numerical solution; black dashed curve:
the polynomial approximation (\ref{midv},\ref{midf}).
}
\label{around-f-zero}
\end{figure}

One can obtain a better approximation of the solution near the zero crossing
by using a polynomial expansion.
The free parameters in this expansion are $\varphi_0$, the position of the zero of $f$,
and $v_0=v(\varphi_0)$.
If one sets $f(\varphi_0)=0$ in (\ref{vdot3}) and (\ref{fdot3}), one obtains $v'(\varphi_0)$
and $f'(\varphi_0)$ as functions of $\varphi_0$ and $v_0$.
Taking one derivative of (\ref{vdot3}) and (\ref{fdot3}) one determines $v''(\varphi_0)$
and $f''(\varphi_0)$ as functions of $\varphi_0$ and $v_0$, and so on.
The resulting expansions read
\footnote{in actual calculations we have pushed this to order $(\varphi-\varphi_0)^{10}$.}
\bea
\label{midv}
v(\varphi)&=&
v_0+0.03377(177.65 v_0-1)\frac{(\varphi-\varphi_0)}{\varphi_0}
+0.07810(177.65 v_0-1)\frac{(\varphi-\varphi_0)^2}{\varphi_0^2}
+\ldots
\\
\label{midf}
f(\varphi)&=&
-0.10414\frac{(\varphi-\varphi_0)}{\varphi_0}
+(-0.05548+1.300v_0+0.0187\varphi_0^2)\frac{(\varphi-\varphi_0)^2}{\varphi _0^2}+\ldots
\eea
\begin{figure}
\includegraphics[width=7.5cm]{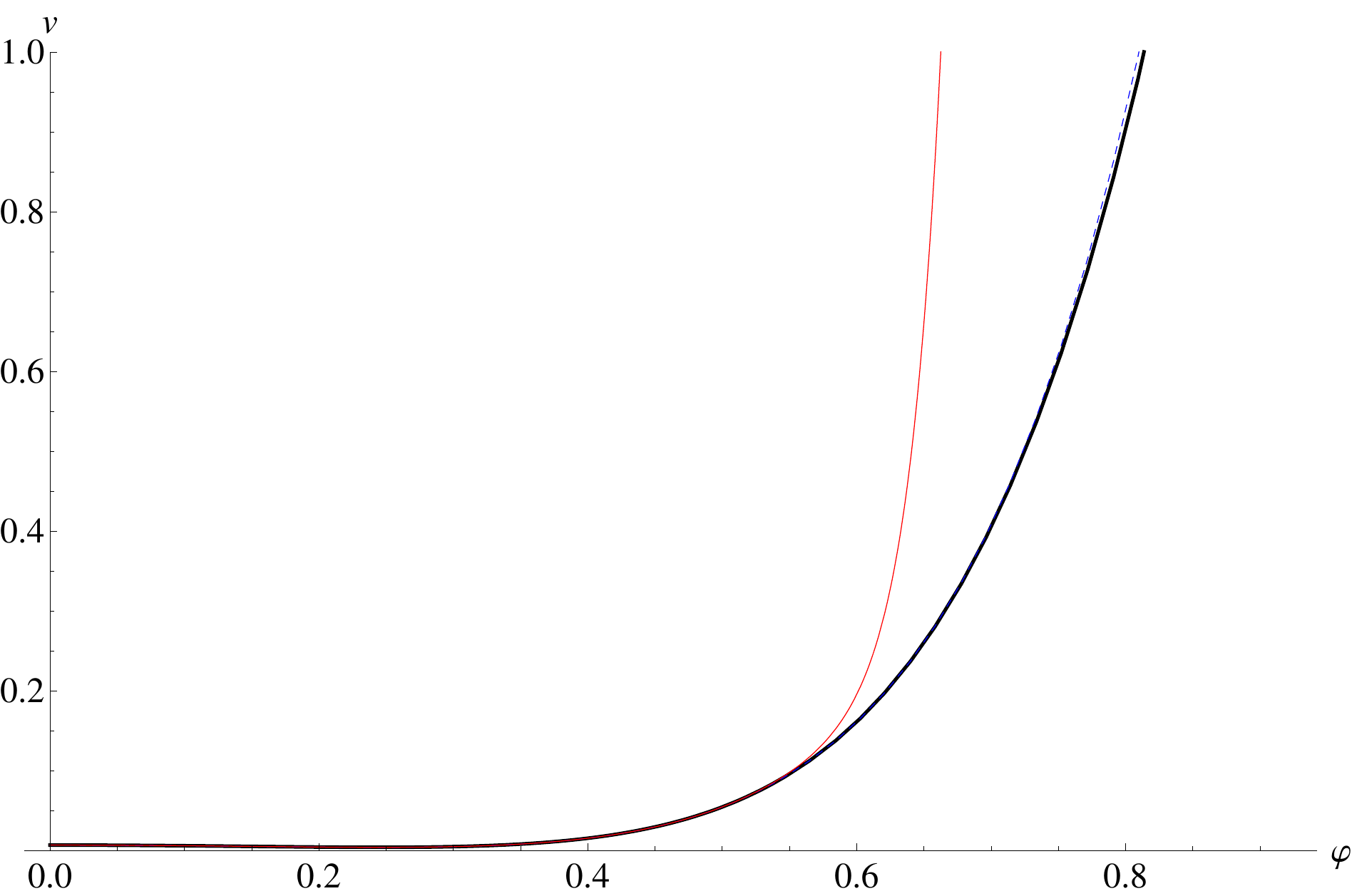}
\
\includegraphics[width=7.5cm]{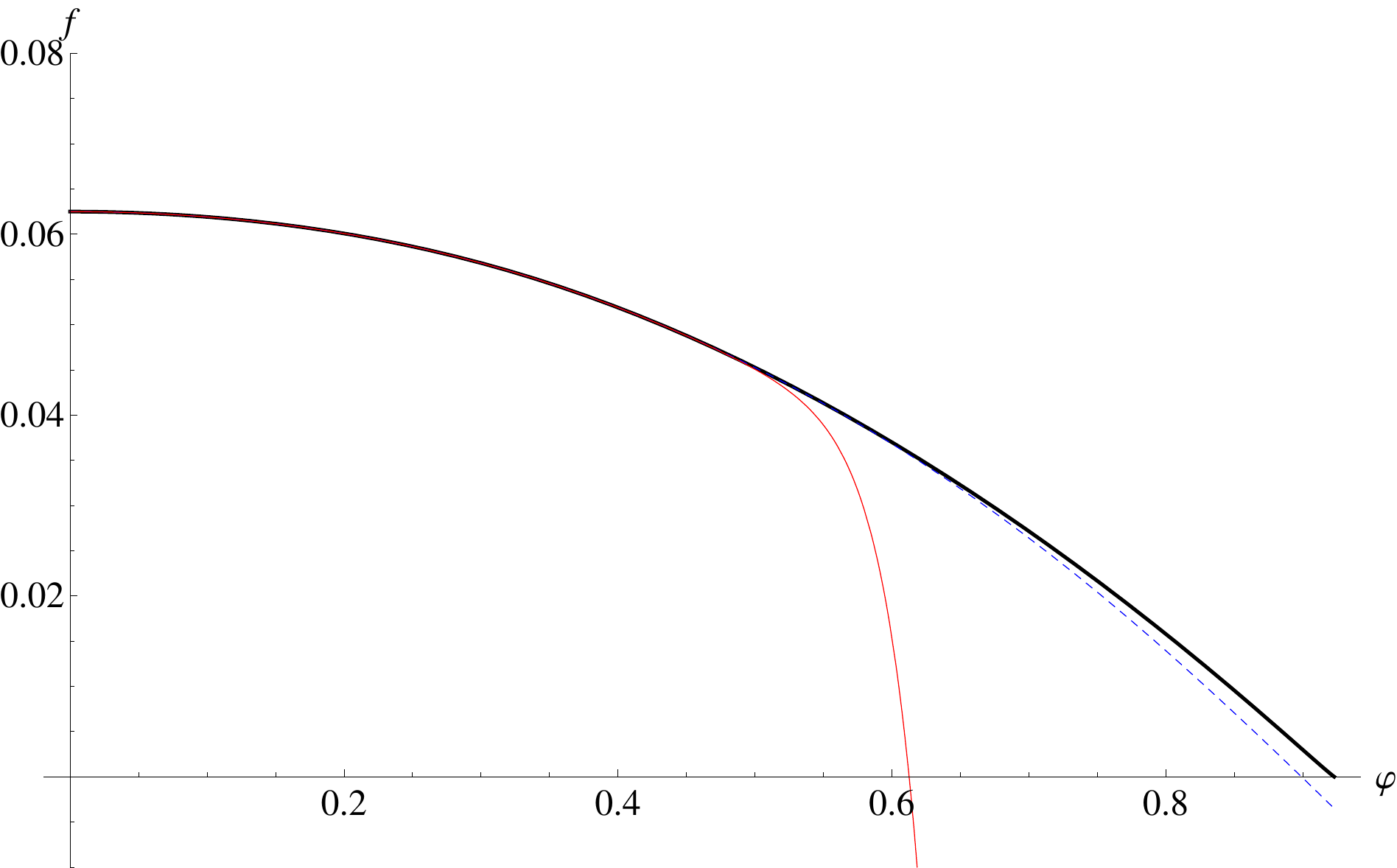}
\caption{The solution up to $\varphi_0$. Left: $v$, right: $f$. Red, continuous curves: the polynomial approximation (\ref{vmin},\ref{fmin});
blue, dashed curves: the Pad\'e approximant;
thick black, continuous curves: the numerical solution.  
The potential has a minimum at $\varphi\approx 0.245$.
}
\label{solnum}
\end{figure}
We then integrated numerically the fixed point equations
from $\varphi_0$ towards lower and higher values of $\varphi$.
In the direction of decreasing $\varphi$ this can be done all the way to $\varphi=0$ 
without undue difficulties.
One then fixes uniquely the initial conditions $\varphi_0$ and $v_0$ 
by demanding that $v'(0)=0$ and $f'(0)=0$.
The resulting numerical solution agrees to high precision with the
numerical solution determined by the spike plot and also with the polynomial solution.
These solutions are shown in Fig.~\ref{solnum}.
The behavior in the region around $\varphi_0$ is enlarged in Fig.\ref{around-f-zero}.
The initial conditions corresponding to the fixed point are
$\varphi_0=0.92299$ and $v_0=2.12474$.
We thus have a reliable numerical solution all the way from $\varphi=0$ to $\varphi=\varphi_0$.
This numerical solution can be continued up to $\varphi\approx35$, but it is not
at all clear that its behavior reflects the actual solution.
In the region $1<\varphi<35$, $v''$ reaches a minimum very close to $-1$ and $f''$ 
passes through a zero. Similar clearly unphysical behavior is also observed 
in the numerical solution of (\ref{vdotscalar}), so we tend to discount
it as due to a numerical instability.
From scalar theory we also know that it is impossible to numerically integrate
the fixed point equation from small to large $\varphi$.
One possibility would be to match the asymptotic large-$\varphi$ behavior to the
expansion around $\varphi_0$.
If one tries to do that, one gets a reasonably good match for $v$
but the match for $f'$ gets no better than a few percent. 
This is because the radius of convergence of the expansion around $\varphi_0$ is zero,
as one can see by analyzing the coefficients of the expansion,
and the asymptotic expansion is not good near $\varphi_0$.
The better procedure would be to start at some large $\varphi$ with initial values dictated
by the asymptotic behavior (\ref{asyv},\ref{asyf})
and to numerically integrate the fixed point equation from large to small $\varphi$.
So far our efforts to do this have been unsuccessful.

\medskip

\section{Scaling solutions in $d=4$}

In this section we look for solutions of the system (\ref{vdot4},\ref{fdot4}).
First we find that also in four dimensions there is a fixed point with constant
$v$ and $f$, which we call again FP1, at
\be
v_*=\frac{3}{128\pi^2}\approx 0.00237\ ;
\qquad
f_*=\frac{41}{768\pi^2}\approx 0.00541\ .
\ee
or equivalently at
\be
\tilde\Lambda_*=\frac{9}{41}\approx 0.219\ ;
\qquad
\tilde G_*=\frac{48\pi}{41}\approx 3.678\ .
\ee

This scaling solution looks like a fixed point of an Einstein-Hilbert theory. If analysed in such a two dimensional subspace, then
the critical exponents are real and simply given by $\theta_1=4$ and $\theta_2=2$ with the corresponding eigenvectors in terms of $(\Lambda,G)$ components
$w_1=(1,0)$ and $w_2 =(\frac{3}{16\pi},1)$.

In terms of $\Lambda$ and $G$ there is also the Gaussian fixed point at $(0,0)$ with critical exponents 
$\theta_1=2$ and $\theta_2=-2$ and the same eigenvectors $w_1=(1,0)$ and $w_2 =(\frac{3}{16\pi},1)$.
Actually the equations are so simple that one can solve for the RG trajectories even
far from the fixed points:
\bea
\tG(t)&=&\frac{\tG(t_0) \tG_*}{\tG(t_0) +\left(\tG_*-\tG(t_0)\right) e^{2(t_0-t)}}\nonumber\\
\tL(t)&=&\tL_* \frac{ \tG(t_0)+\left(\frac{16\pi}{3}\tL(t_0)-\tG(t_0)\right)e^{4(t_0-t)}}{\tG(t_0) +\left(\tG_*-\tG(t_0)\right) e^{2(t_0-t)}}\ .
\eea
Complete RG trajectories (with $-\infty<t<\infty$) exist for initial conditions 
such that $\tG(t_0)< \tG_*$, otherwise a Landau pole is present.
We note that the separatrix which connects the Gaussian fixed point to FP1 is a straight segment, along the direction of $w_2$.
One can see that the corresponding dimensionful couplings $G\sim\tG e^{2(t_0-t)}$ and $\Lambda\sim\tL e^{-2(t_0-t)}$ for $\tG(t_0)< \tG_*$ reach a constant value in the IR.

Let us recall that the FP1 solves the fixed point equations (\ref{vdot4},\ref{fdot4}).
It is therefore important to study the eigenperturbations in such a space.
For this particular solution the linearization leads to the equations
\bea
&{}&-(\lambda +4)\delta v+\varphi \, \delta v'-\frac{\delta v''}{32 \pi ^2}=0 \nonumber\\
&{}&-(\lambda +2) \delta \!f+\varphi \, \delta \! f'-\frac{\delta \!f''}{32 \pi ^2}+\frac{\delta v''}{96 \pi ^2}=0
\label{eigeneq1}
\eea
whose solutions can be investigated both analytically and numerically.
Imposing that the eigenfunctions are even functions of $\varphi$ leads to a quantization condition for the eigenvalues: $\lambda=-\theta=-4,-2,0, ...$
(with $\theta=-\lambda$ the critical exponent).
We find
\bea
&{}&\theta_1=4 , \quad \quad w_1^t=(\delta v,\delta\! f)_1 = (1,0) \nonumber\\
&{}&\theta_2=2 , \quad\quad w_2^t=(\delta v,\delta\! f)_2 = (0,1) \nonumber\\
&{}&\theta_3=0 , \quad\quad w_3^t=(\delta v,\delta\! f)_3 = \left(0,-\frac{1}{32\pi^2}+\varphi^2\right) \ .
\eea
The dimension of the UV critical surface is $2$ without taking into account the marginal direction, whose behavior should be further analyzed.

Since the relevant directions of this fixed point are related to scalar field-independent eigenperturbations, 
if the bare action in the UV is located in the functional space close to it, the flow towards the IR may tend to amplify the constant components of $v$ and $f$. 
In such a case quantum effects could lead to a reduced field dependence in the effective action, 
where all quantum fluctuations are integrated out. 
Note that for effective actions with a reduced $\varphi$ dependence the equations of motions
in such a model would have solutions closer to be maximally symmetric spaces, e.g. de Sitter geometry.

There is then a nontrivial analytic scaling solution FP2 with constant $v$ but non-constant $f$:
\be
v_*=\frac{3}{128\pi^2}\approx 0.00237\ ,\qquad 
f_*(\varphi)=\frac{37}{768\pi^2}+\frac{1}{6}\varphi^2
\approx  0.00488+0.167\varphi^2\ ,
\ee
which corresponds to
\be
\tilde G_*=\frac{48\pi}{37}\approx 4.08\ ;\qquad 
\tilde\Lambda_*=\frac{9}{37}\approx0.243\ ;\qquad
\xi=\frac{1}{6}\,.
\ee

The linearized equations for the eigenperturbations around this fixed point read:
\bea
&{}&- (\lambda +4)\delta v+\varphi \delta v'-\frac{\left(128 \pi ^2 \varphi ^2+37\right)}{32 \pi ^2 \left(384 \pi ^2 \varphi ^2+37\right)} \delta v''=0 
\nonumber\\
&{}&- (\lambda +2)\delta \!f+\varphi  \delta\! f'-\frac{\left(128 \pi ^2 \varphi ^2+37\right)}{32 \pi ^2 \left(384 \pi ^2 \varphi
   ^2+37\right)} \left(\delta \!f''-\delta v''\right)=0\ .
\label{eigeneq1}
\eea

Also here, as in the case with $d=3$, we proceed with numerical methods.,
which give a good approximation for the eigenfunctions for small values of $\varphi$.
We find three relevant direction with critical exponents: $\theta_1=4$, $\theta_2=2$, $\theta_3=1.809$. In particular
\bea
&{}&\theta_1=4 , \quad \quad  \ \ w_1^t=(\delta v,\delta\! f)_1 = (1,0) \nonumber\\
&{}&\theta_2=2 , \quad\quad  \ \ w_2^t=(\delta v,\delta\! f)_2 = (0,1) \nonumber\\
&{}&\theta_3=1.809 , \quad w_3^t=(\delta v,\delta\! f)_3 = (c_{1v}+\varphi^2,c_{1f}+\varphi^2) 
\eea
where $c_{1v}\simeq -0.0028$ ad $c_{1f}\simeq 0.53$.

Starting from an initial condition close to this scaling solution a flow towards the infrared will enhance such components of $v$ and $f$ in the effective action
and in general will develop also a non trivial potential.
It is not clear if along the flow towards the IR there is a possibility for the potential to undergo a dynamical quantum symmetry breaking.
We shall return to these points elsewhere.

There is also the analog of the fixed point FP3 that we found in three dimensions.
It occurs at 
\be
\label{other}
v_*=\frac{3}{128\pi^2}\approx 0.002374\ ;\qquad 
f_*=-\frac{41}{420}\varphi^2\approx -0.09762 \varphi^2\ ,
\ee
It has the same unappealing properties as in three dimensions and will not be discussed further in this section.

\begin{figure}
\includegraphics[width=12cm]{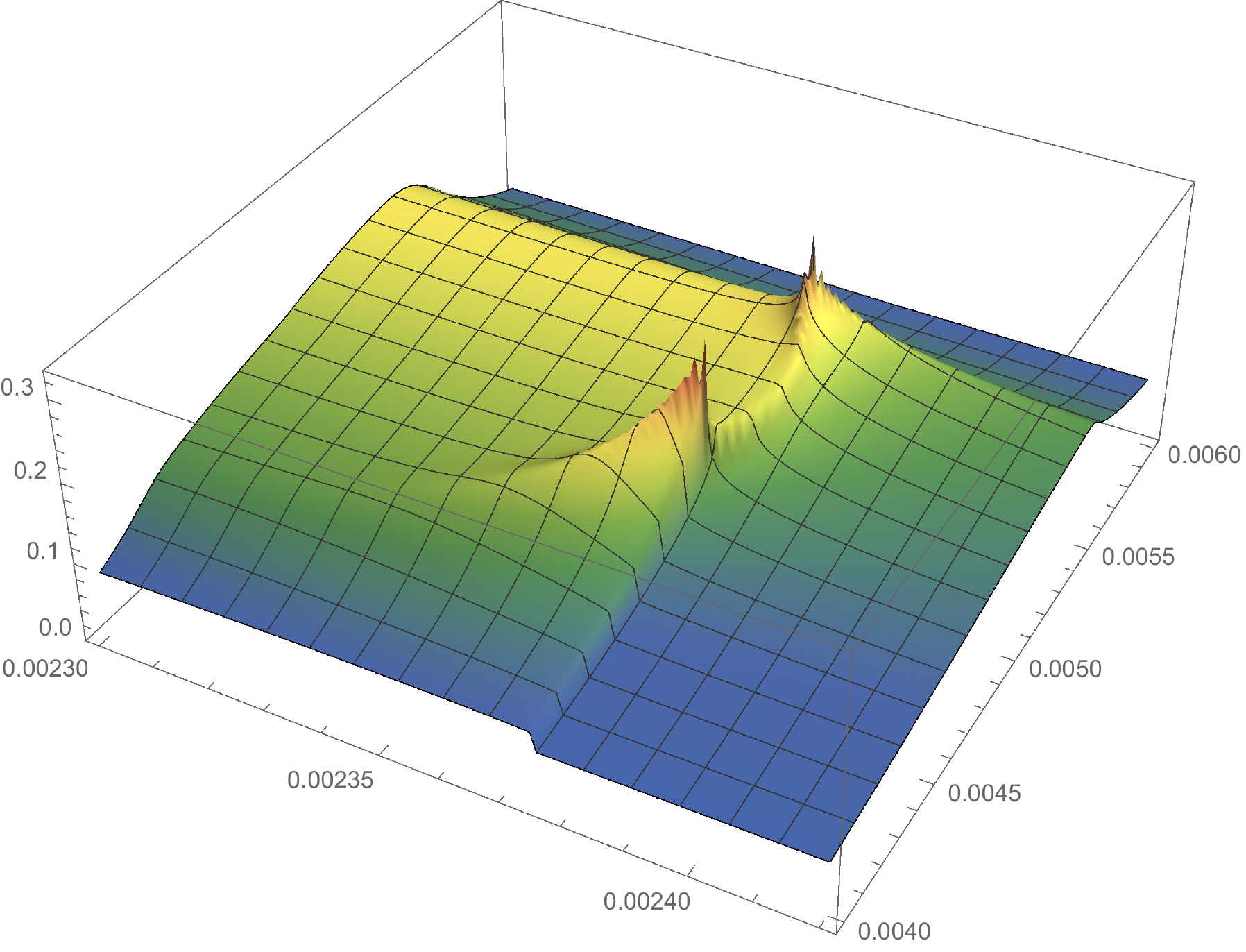}
\caption{Plot of the maximum value of $\varphi$ reached by the numerical integrator depending of the initial conditions $0.0023<v(0)<0.00242$ and $0.004<f(0)<0.006$ .}
\label{spike1_d4_v0_f0}
\end{figure}

Finally we would like to investigate the fixed point equations in order to search for other non trivial solutions, possibly with a non trivial scalar potential.
We have used all the methods already described in the three-dimensional case.
Finite polynomial approximations do not show solutions that persist systematically
from one truncation to the next, except for FP1 and FP2.
We have also searched for numerical solutions starting from $\varphi=0$,
with the initial conditions $v'(0)=f'(0)=0$.
We summarise our findings in a three dimensional spike plot in Fig~\ref{spike1_d4_v0_f0}.
The two spikes that are visible correspond to FP1 and FP2. 
As noted in footnote 8, FP3 cannot be seen in this way.
No other spikes have been found which lead to a nontrivial potential that is bounded from below.

We complement this information with an asymptotic analysis. After determining the possible asymptotic behaviors of the solutions of the fixed point equations (\ref{vdot4},\ref{fdot4}),
we try to construct them integrating numerically from compatible initial conditions in the asymptotic region towards the origin.
The first four terms of the asymptotic behavior which starts with the classical scaling read
\bea
\label{asyv4}
v(\varphi)&=&A \varphi^4+\frac{1}{64\pi^2}
+\frac{1+12B}{2304A\pi^2 \varphi^2 }
-\frac{135A+16\pi^2(1+12B)^2}{589824 \pi^4 A^2 \varphi^4}
+\ldots
\\
\label{asyf4}
f(\varphi)&=& B\varphi^2+\frac{15}{256\pi^2}
-\frac{1+12 B}{4608\pi^2 A\varphi^2}
+\frac{135A+16\pi^2(1+6B-72B^2)}{1327104 \pi^4 A^2 \varphi^4}
+\ldots
\eea

We have not been able to find new non trivial numerical solutions starting from large values of $\varphi$, but we do find the solutions already known analytically.
In doing this we have used asymptotic expansion with up to 20 terms for each function.

We have also tried to investigate solutions for which there exist a value $\varphi_0$ where $f(\varphi_0)=0$ by analyzing them in terms of a polynomial expansion, similarly to the $d=3$ case.
The two parameters which parametrize the solutions are again $\varphi_0$ and $v(\varphi_0)$.
Trying to fix them by numerically evolving towards the origin we find that the solution such that $v'(0)=f'(0)=0$ can be reached only for $\varphi_0 \to \infty$,
in which case one reproduces FP1 for small $\varphi$ values. No nontrivial solution with $f$ changing sign seems to exist.
Thus all methods point to the same conclusion, namely that there are no global scaling solutions
in $d+4$ beyond the ones we have found in closed form.

\section{Some results keeping the $\dot F$ dependence on the r.h.s.}

We discuss here the flow equations keeping the terms proportional to $\dot F$ in the r.h.s..
We proceed in a way similar to the previous analysis of the last two sections, considering the two cases in $d=3$ and $d=4$ dimensions. 
We find two analytic scaling solutions,
which correspond clearly to the solutions FP1 and FP3 of the previous sections,
but the solution FP2 is not present.
In appendix $C$ we give the general results for $d$ dimensions for the fixed point equations,
their two sets of analytic solutions and the flow equations.

\subsection{$d=3$}

The fixed point equation for $d=3$ reads
\bea
\label{vdot3full}
0&=&
-3\,v+\frac{1}{2}\varphi\,v'+\frac{4}{15\pi^2}-\frac{\varphi  f'}{30 \pi ^2 f}-\frac{\varphi  f' \left(8 f''+v''+1\right)+10 f v''}{60 \pi ^2 \left(4 f'{}^2+f \left(1+v''\right)\right)}\\
\label{fdot3full}
0&=&
-f+\frac{1}{2}\varphi\,f'+\frac{67}{90\pi ^2}-\frac{29 \varphi  f'}{360 \pi ^2 f}+
\frac{\varphi  f' \left(8 f''+v''+1\right)+6 f v''}{144 \pi ^2 \left(4 f'{}^2+f (1+v'')\right)}-\nonumber\\
&{}&-\frac{\left(f f''+2 f'{}^2\right) \left(5 f^2+2 \varphi  f'{}^3+4 f f' \left(5 f'-\varphi  f''\right)\right)}{30 \pi ^2 f \left(4
   \left(f'\right)^2+f \left(1+v''\right)\right)^2}\eea

There is again a solution with constant $v$ and $f$:
\be
v_*=\frac{4}{45\pi^2}\approx 0.009006\ ;
\qquad
f_*=\frac{67}{90\pi^2}\approx 0.07543 \nonumber\\
\ee
Although slightly shifted, this is clearly the analog of FP1 and we shall call it again by the same name.
The linearized flow equations at FP1 read
\bea
0 &=&-(\lambda+3) \delta v+\frac{1}{2}\varphi  \delta v'+\frac{9}{67} \left( \lambda \delta \! f-\frac{1}{2} \varphi \delta \! f'\right)
 -\frac{\delta v''} {6 \pi ^2} \nonumber\\
0 &=& -\left(\frac{215} {268} \lambda+1\right) \delta \! f+\frac{215}{536} \varphi  \delta \! f'
-\frac{\delta \!f''}{6 \pi ^2}+\frac{\delta v''}{24 \pi ^2}
\eea

Studying  these equations we find that the scaling solution FP1 now 
has five relevant directions and one marginal,
with the following approximate critical exponents and eigenperturbations:
\bea
&{}&\theta_1=3 , 
\qquad \qquad   w_1^t
=(\delta v,\delta\! f)_1 = (1,0) \nonumber\\
&{}&\theta_2=2 , 
\qquad\qquad   w_2^t=(\delta v,\delta\! f)_2 
= \left(-\frac{8}{27 \pi ^2}+ \varphi^2, -\frac{67}{486 \pi ^2}\right) \nonumber\\
&{}&\theta_3=\frac{268}{215} , \quad  \quad \ \ \, w_3^t
=(\delta v,\delta\! f)_3 
= \left(-\frac{36}{377}, 1\right)\\
&{}&\theta_4=1 , 
\qquad  \qquad w_4^t=
(\delta v,\delta\! f)_4 
=  \left(\frac{362}{1431 \pi ^4}-\frac{16}{9 \pi ^2} \varphi^2+\varphi^4, \frac{8308}{12879 \pi ^4} -\frac{67}{81 \pi ^2}\, \varphi^2\right) 
\nonumber\\
&{}&\theta_5=\frac{53}{215} , 
\quad  \quad \ \ \, w_5^t
=(\delta v,\delta\! f)_5  
= \left(\frac{200049}{11996140 \pi ^2}-\frac{36}{377} \varphi^2, -\frac{17741}{350304 \pi ^2} -\frac{68}{123} \varphi^2\right)\nonumber\\
&{}&\theta_6=0 , \qquad  \qquad w_6^t
=(\delta v,\delta\! f)_6
=\bigg(-\frac{1810}{4293 \pi ^6}+\frac{1810 }{477 \pi ^4}\varphi ^2-\frac{40}{9 \pi ^2} \varphi ^4+\varphi ^6,
\nonumber\\
&{}&
\qquad\qquad
\qquad\qquad
\qquad\qquad
\qquad\qquad
\qquad
-\frac{74935}{25758 \pi ^6}+\frac{41540 x}{4293 \pi ^4}\varphi ^2-\frac{335 }{162 \pi ^2}\varphi ^4\bigg) \,.
\nonumber
\eea
The two pairs of degenerate solutions with critical exponents 1 and 0 that we found in section IV
have split, with one in each pair becoming slightly more relevant.

There is no solution of (\ref{vdot3full},\ref{fdot3full}) that could be identified with FP2.
The fixed point FP3, however, is still present exactly in the same position:
\be
v_*=\frac{1}{18\pi^2}\ ;
\qquad
f_*(\varphi)=-\frac{43}{568}\varphi^2\ .
\ee
The eigenperturbations around FP3 are different.
We have analyzed them numerically but we do not find it very useful to show them here.

\subsection{$d=4$}

In four dimensions the fixed point equations are
\bea
\label{vdot4full}
0&=&
-4\,v+\varphi\,v'+\frac{5}{32\pi^2}-\frac{5 \varphi  f'}{192 \pi ^2 f}
-\frac{\phi  f' \left(6 f''+v''+1\right)+6 f v''}{192 \pi ^2 \left(3 f'{}^2+f \left(1+v''\right)\right)}
\\
\label{fdot4full}
0&=&
-2f+\varphi\,f'+\frac{157}{1152\pi ^2}-\frac{5 \varphi  f'}{288 \pi ^2 f}+\frac{\varphi  f' \left(6 f''+v''+1\right)+
4 f v''}{384 \pi ^2 \left(3 f'{}^2+f (1+v'')\right)}\nonumber\\
&{}&
-\frac{\left(f f''+f'{}^2\right) \left(2 f^2+\varphi  f'{}^3+6 f f'{}^2-2  \varphi f f' f''\right)}{64 \pi ^2 f \left(3
   f'{}^2+f(1+ v'')\right)^2}
\eea

The analytic solution FP1 is given by
\be
v_*=\frac{5}{128\pi^2}\approx 0.003958\ ;
\qquad
f_*=\frac{157}{2304\pi^2}\approx 0.006904\ . \nonumber
\ee
Linearizing the flow equation given in Eqs.~(\ref{flowvfull},\ref{flowffull}) for $d=4$ around FP1 we find:
\bea
0 &=&-(\lambda +4) \delta v+\varphi  \delta v'+\frac{72}{157} \left(  \lambda \delta \! f -\varphi \delta\!f'\right) -\frac{\delta v''} {32 \pi ^2} \nonumber\\
0 &=& -\left(\frac{123} {157} \lambda+2\right) \delta \! f+\frac{123} {157} \varphi  \delta \! f'
-\frac{ \delta \!f''}{32 \pi ^2}+\frac{\delta v''}{96 \pi ^2}
\eea
%
Studying analytically these equations we find that FP1 has three relevant and one marginal direction:
\bea
&{}&\theta_1=4 , \qquad \qquad  \qquad \ \ \, w_1^t=(\delta v,\delta\! f)_1 = (1,0) \nonumber\\
&{}&\theta_2=\frac{314}{123}\simeq 2.553 , \qquad   w_2^t
=(\delta v,\delta\! f)_2 
= \left(-\frac{72}{89}, 1\right) 
\nonumber\\
&{}&\theta_3=2 , \qquad  \qquad   \qquad \ \ \, w_3^t
=(\delta v,\delta\! f)_3 
= \left(-\frac{29}{544\pi^2}+\varphi^2,\frac{157}{3264\pi^2}\right) \\
&{}&\theta_4=\frac{68}{123}\simeq0.553 , \qquad  w_4^t
=(\delta v,\delta\! f)_4 
= \left(\frac{56913}{3094352 \pi ^2}-\frac{72}{89} \varphi^2,-\frac{17741}{350304 \pi ^2}+\varphi^2\right)
\nonumber\\
&{}&\theta_5=0 , \qquad  \qquad \qquad \ \ \, w_5^t
=(\delta v,\delta\! f)_4 
= \bigg(\frac{87}{17408 \pi ^4}
-\frac{87}{272 \pi ^2}\varphi^2
+\varphi^4,-\frac{215}{17408 \pi ^4}
+\frac{157}{544 \pi ^2}\varphi^2\bigg)\ .
\nonumber
\eea

The fixed point FP2 is absent but FP3 is again present in the same position:
$v_*=\frac{3}{128\pi^2}$, $f_*(\varphi)=-\frac{41}{420}\varphi^2$.
It has four relevant directions, with critical exponents $4$, $2.104$, $1.574$, $0.2475$.
The eigenvector corresponding to eigenvalue $4$ has components $(\delta v,\delta\! f) = (1,0)$,
the others have been studied numerically.
We find that if $\delta v$ is chosen to be increasing for large $\phi$,
as is required by physical considerations, then $\delta f$ is negative
and decreasing for large field.
Given that $f$ is already negative, this is again physically undesirable.

Whether the true scaling solutions of the theory (assuming that any exists) resemble more this case or
the one described in section V, is something that will have to be studied taking into
account the constraints inposed by the split-symmetry Ward identities.

\section{Conclusions}

In this work we have looked at ways to improve the flow equations for the
functions $f$ and $v$ characterizing the EAA for scalar-tensor gravity,
truncated at two derivatives.
We have argued in favor of using the exponential parametrization of the metric fluctuation,
which is more natural if one takes into account the fact that the metric carries a 
nonlinear realization of the linear group,
together with a physical gauge fixing that simply removes the gauge degrees of freedom
from the calculation.

As usual the EAA has been truncated: aside from the gauge fixing, ghost and cutoff terms,
we have assumed that the EAA is a function of the
full metric $g_{\mu\nu}$ and of the full scalar field $\phi$, 
and not of $\bar g_{\mu\nu}$ and $h_{\mu\nu}$, $\bar\phi$ and $\delta\phi$ separately.
This is known as a ``single field approximation'' and its limitations have come 
under close scrutiny recently \cite{mr,donkin,becker,dietz3}.
Within the single metric approximation but in contrast to much of the previous
literature, here we have retained infinitely many couplings,
namely those pertaining to the first order of the derivative expansion in the metric,
and the local potential approximation in the scalar.
This expansion is well motivated in applications to critical phenomena,
less so when studying the UV limit, but this approach is still more systematic than
additionally truncating on powers of the scalar field.
The consistency of the truncation should be tested by
enlarging it: in the scalar sector the next step would be
to take into account the wave function renormalization of $\phi$,
while in the gravity sector it would consist of
including all four-derivative terms, or perhaps a truncation of the
form $F(R,\phi)$, as discussed earlier in \cite{narainrahmede}.

We have considered two approximation schemes, one where we dropped terms proportional 
to $\dot F$ in the r.h.s. of the equations and a second one where we keep them.
This is similar to, but obviously more general than, neglecting $\dot G$ 
on the r.h.s. of the flow equations for the Einstein-Hilbert truncation.
Let us stress that the anomalous dimension of the quantum fluctuations of the metric
is not directly related to $f$.
An approximate computation of the anomalous dimension can be obtained from the flow 
of the two point function of the fluctuation field, along the lines of~\cite{codello,dona3}.

The new equations do not have a singularity for $v=f$, as previous equations did.
One consequence is that in the two dimensional Einstein-Hilbert subspace 
(where $v$ and $f$ are constant) there is
no singularity for $\tilde\Lambda=1/2$, nor for any other value of $\tilde\Lambda$.
We expect that this feature will be useful to study the IR limit of the theory.
Another consequence is that in three dimensions
the new equations admit, at least neglecting $\dot F$ and for sufficiently small fields, 
a nontrivial fixed point that resembles very closely the Wilson-Fisher fixed point of scalar theory
in the LPA approximation.
We have traced this similarity to the fact that the flow equation for $v$, eq.(\ref{vdot3}),
reduces, for constant $f$, to the flow equation of scalar theory (\ref{vdotscalar}).
Recall that, in contrast, the ``old'' equations admitted only solutions with negative Taylor coefficients \cite{narain}.
The function $f$ starts positive at $\varphi=0$ and decreases monotonically.
The solution exists at least up to the point $\varphi_0$ where $f$ crosses zero.
At this point the new equations have a singularity,
but we have shown by analytic methods that the singularity can be traversed.
We have not been able so far to join the expansion around $\varphi_0$
to the expansion around infinity.
This may be a technical issue that could be overcome by
using more powerful numerical techniques,
but there may also be deeper issues due to the vanishing of many terms in the Hessian
when $F=0$.

In addition to this tentative gravitationally dressed Wilson-Fisher
fixed point in three dimensions, we have also shown that, in any dimension,
the equations where one neglects the dependence of $\dot F$ in the r.h.s.
admit three other simple fixed points:
a ``Gaussian Matter Fixed Point'' FP1 with constant $v$ and $f$,
a fixed point FP2 with $v$ constant and $f=f_0+f_1\varphi^2$, with $f_0$ and $f_1$ positive
(in the interesting range of dimensions)
and a fixed point FP3 with $v$ constant and $f=f_1\varphi^2$, with $f_1$ negative.
We emphasize that although these fixed points have only a few nonvanishing couplings,
they are not just the solution of a polynomial truncation
but of the full fixed point equations for the functions $v$ and $f$.
In three dimensions FP2 is characterised by four relevant eigenperturbations
while FP1 has four relevant and two marginal eigenperturbations.
In four dimensions we have shown that FP1 and FP2 have only three relevant
(or marginal) deformations, making them good candidates for asymptotically safe models. 
We also note that the critical exponents we find are not related to redundant operators,
in contrast to what was seen for the flow equation of $f(R)$ gravity in \cite{dietz2}.
In any dimension, the fixed points FP1 and FP3 are also present, with little or no
change in their properties, when the terms proportional to $\dot F$ are retained
in the r.h.s..
Unfortunately the physically interesting fixed point FP2 disappears in this case.
We do not know at present whether this reflects a genuine property of the underlying physics.

One may be worried by this and more generally by the different properties 
of the equations that we have found here,
compared e.g. to those of \cite{narain} or to the flow equation in the physical gauge
$\xi'_\mu=\sigma'=0$ (as discussed in the end of section III.B).
The situation is similar to that of $f(R)$ gravity, where different equations
turned out to have rather different solution spaces \cite{dietz1}.
The answer seems to be that some approximations are too drastic:
all equations are good enough to find the fixed point
within finite dimensional truncations, but the study of its properties in
an infinite dimensional function space is more delicate and requires better approximations.
It has been shown in \cite{dietz3} that pathological features resembling those 
encountered in $f(R)$ gravity can be artificially induced even in pure scalar theory
by an improper use of the background field method.
In particular, one should pay close attention to the violation of split symmetry 
(\ref{splitsymmetry}), which, at linear level, amounts to
$\delta \phi \to \delta \phi+\delta \psi$, $\bphi \to \bphi -\delta \psi$ in the scalar sector and
$h_{\mu\nu}\to h_{\mu\nu}+\delta h_{\mu\nu}$, $\omega\to\omega+\delta\omega$,
$\bar g_{\mu\nu}\to\bar g_{\mu\nu}-\delta h_{\mu\nu}-2\bar g_{\mu\nu}\delta\omega$ in the gravitational sector.
While an investigation of this point will be necessary,
it seems that the equations derived here are already powerful enough to
discover at least some of the scaling solutions in the theory.
This may be a hint that, within the single-field approximation, 
the use of the exponential parameterization 
and of the unimodular physical gauge is to be preferred.

Let us note that unimodular gravity corresponds to the case where the conformal fluctuations $h$  are completely absent.
For such a theory, in the single field approximation for the average effective action, the flow equation is obtained
from the one of full gravity in the unimodular gauge by removing the corresponding ghost contribution, 
which is a constant term in both equations for $v$ and $f$. 
Since the running of $f$ does not depend on $v$ but only on its derivatives, this is consistent
with the fact that in unimodular gravity the constant term in the potential is an integration constant. 
In this framework any constant value of $v$ at the fixed point would not contain any physical information.

There are several obvious extensions of the truncation that we plan to return to in the future.
Also, we have focused here mainly on the mathematical properties of the system
of flow equations, but ultimately one is interested in physical applications.
In this regard we observe that the fixed point FP2 in $d=4$ has the properties that were discussed
in \cite{hprw} as prerequisites for the construction of interesting cosmological models.
With the linearized perturbations given here and with numerical integration of the flow equation
it will be possible to analyze in detail several scenarios.

{\it Note added}. While this paper was being considered for publication, a work by Borchardt and Knorr \cite{bk} appeared
where they describe the use of pseudo-spectral methods
to solve linear and non-linear ODEs.
They describe a numerical solution 
of the full fixed point equations in three dimensions
(\ref{vdot3full},\ref{fdot3full}).
In contrast to the solution of the equations
(\ref{vdot3},\ref{fdot3}) that we described in section IV.B,
this solution has a growing $f$ that never crosses zero.

\appendix

\section{Multiplicative quantum-background split}

The metric is generally thought of as a tensor, hence as a ``linear'' object.
This can be misleading. In General Relativity the metric is subject to the constraint
of being non-degenerate and of having a fixed signature. These constraints
define a nonlinear subspace in the space of symmetric tensors, and the metric
is assumed to lie in this subspace.
To understand this subspace better, observe that a metric can be defined
by giving a frame and declaring it to be orthonormal.
This defines a surjective map from the set of all frames, 
which is in one-to-one correspondence with the group $GL(d)$, to the set of metrics.
Two frames that differ by the action of a (pseudo)-orthogonal transformation
define the same metric. Hence the set of all metrics with signature $(p,q)$
at a point can be identified with the coset space $GL(d)/O(p,q)$,
which is an open subset in the space of symmetric tensors.
It is not a priori clear whether the gravitational path integral
should be restricted to this subset or could include also degenerate
metrics or even metrics with other signatures \cite{per1}.
We assume here the former point of view.
Then, it is not clear, when one employs the standard linear background-fluctuation split
\be
\label{linsplit}
g_{\mu\nu}=\bar g_{\mu\nu}+h_{\mu\nu}\ ,
\ee
how to implement the requirement that the quantum fluctuations
should stay within the space of metrics of a given signature.

From this point of view it is more natural to use a multiplicative version of
the background field method, introduced previously in (\cite{flop}).
\footnote{There, we were only interested in a perturbative evaluation of
the path integral and therefore $\theta$ was subsequently expanded
additively in small fluctuations around a background $\bar\theta$.
Here we shall not do this.}
Any metric of given signature in $d$ dimensions can be written as
\be
\label{mbg}
g_{\mu\nu}=\theta^\rho{}_\mu\theta^\sigma{}_\nu \bg_{\rho\sigma}
\ee
where $\bg$ is a fixed metric of the same signature as $g$
and $\theta^\rho{}_\mu$ is a field with values in the group $GL(d)$.
The multiplicative quantum-background split (\ref{mbg}) has an inherent
arbitrariness that consists of local $GL(d)$-transformations
\be
\label{splitsymmetry}
\theta'^\alpha{}_\mu(x)=\Lambda^{-1}(x)^\alpha{}_\beta\theta^\beta{}_\mu(x)\ ;\qquad
\bg'_{\rho\sigma}(x)=\Lambda^\alpha{}_\rho(x)\Lambda^\beta{}_\sigma(x)\bg_{\alpha\beta}(x)\ .
\ee
This is a multiplicative version of the ``split symmetry''
$\bar g_{\mu\nu}\to \bar g_{\mu\nu}+\epsilon_{\mu\nu}$,
$h_{\mu\nu}\to h_{\mu\nu}-\epsilon_{\mu\nu}$
of the linear split (\ref{linsplit})
and is mathematically similar to a gauge invariance.
Note that when $\bar g_{\mu\nu}=\delta_{\mu\nu}$ the field $\theta$ 
can be viewed as an orthonormal frame. 
(We discuss here the Euclidean case.
Aside from global issues,
other signatures can be treated in the same way).
Every $\bar g_{\mu\nu}$ can be regarded as a split-symmetry-transform of $\delta_{\mu\nu}$,
so there is a subgroup of $GL(d)$ that is conjugate to $O(d)$ and leaves $\bar g$ invariant.
We shall refer to it as the stabilizer of $\bar g$.
In a neighborhood of the identity, one can write $\theta=e^X$, where $X^\rho{}_\mu$ is a generic matrix.
Let $Y_{\mu\nu}=\bg_{\mu\rho}X^\rho{}_\nu$.
For any $g$ and $\bar g$, it is possible, by means of a split-symmetry transformation
belonging to the stabilizer of $\bar g$, 
to choose the matrix $X$ in such a way that $Y$ is symmetric.
Then it is easy to check that
\bea
g=e^{X^T}\bg e^X
=e^{Y\bg^{-1}}\bg\, e^{\bg^{-1}Y}
=\bg e^{\bg^{-1}Y} e^{\bg^{-1}Y}
=\bg\, e^{2X}\ .
\eea
Putting $2X=h$, this is our starting point (\ref{decomp}).

\section{Jacobian for the functional measure}

In this paper we have assumed that the functional integral over metrics has
a simple measure when the field $h_{\mu\nu}$ is defined by the exponential parametrization.
An alternative point of view is that the measure is simple when $h_{\mu\nu}$
is defined by the usual linear parametrization.
In this case there would be a nontrivial Jacobian, which we calculate in this Appendix and, 
for the truncation we have chosen, does not contribute to the flow equations.

On differentiating the matrix $g= \bar{g} \,e^h$ we obtain at each point of space-time
\be
{\rm d} g = \bar{g} \sum_{n=0}^\infty \frac{1}{n!} {\rm d} (h^n)
\ee
where
\be
{\rm d} (h^n) = \sum_{k=0}^{n-1} h^k ({\rm d} h) h^{n-1-k}=\sum_{k=0}^{n-1} \left[ h^k \otimes \left(h^T\right)^{n-1-k} \right] dh
\ee
The last equality can be understood easily if we make explicit use of indices (repeated indices are implicitly summed):
\be
(X dh Y)_{ij}=X_{ia} dh_{ab} Y_{bj}=X_{ia} Y^T_{jb} dh_{ab} = \left[ X \otimes Y^T \right]_{ij,ab} dh_{ab}= \left( \left[ X \otimes Y^T \right] dh \right)_{ij}
\ee
Therefore we have for the functional integration measure
\be
[{\rm d} g] = {\rm Det}[\bar{g} M] [d h]
\ee
where 
\be
M(x)=\sum_{n=0}^\infty \frac{1}{n!}  \sum_{k=0}^{n-1}\left[ h^k \otimes \left(h^T\right)^{n-1-k} \right] \,.
\ee

We can consider a representation of the determinant based on functional integration over Grassmann fields:
\be
{\rm Det}[M] = \int [{\rm d} \theta {\rm d} \bar{\theta}] e^{\int \bar{\theta}(x) M(x) \theta(x)}
\label{extra_action}
\ee

We note that within the Functional RG approach we can employ a coarse-graining also on the $\theta$ and $\bar{\theta}$ fields
which can be considered as quantum fluctuations. One can therefore see that in the background field approach for a single metric
truncation the flow is not affected by the extra term in the action given by (\ref{extra_action}). 
Indeed if $h \ne 0$ this is an interaction term which we may choose to not include in the lowest order truncation. 
This measure would affect the computation of the anomalous dimension of the quantum metric.
Moreover it would come into play when adopting more complicated (bimetric) truncations.

Alternatively one can compute directly the determinant of $M(x)$ at each point of space-time if the matrix $h$ can be diagonalized
by a similarity transformation. Unfortunately in general $h$ is not symmetric since $h=\bg^{-1} Y$ where $Y$ is symmetric. 
Let us note that for the case of interest to us, like a maximally symmetric Euclidean background, 
we can eventually choose a coordinate system for which $h$ is symmetric (gauge fixed background).
On general grounds if we can diagonalise $h$ by a similarity transformation
so that $h=R \Lambda R^{-1}$ with $\Lambda={\rm diag}\{\lambda_i\}$ a diagonal matrix, then $M(x)$ is similar to the diagonal matrix
\be
D(x)=\sum_{n=0}^\infty \frac{1}{n!} \sum_{k=0}^{n-1} \left[ \Lambda^k \otimes \Lambda^{n-1-k} \right] 
\ee
which has the same determinant of $M(x)$. One finds
\bea
{\rm det} M(x)&=&\prod_{ij}  \sum_{n=0}^\infty \frac{1}{n!} \sum_{k=0}^{n-1}\lambda_i^k  \lambda_j^{n-1-k} 
= \prod_{i}  \sum_{n=0}^\infty \frac{1}{n!} (n \lambda_i^{n-1})+
\prod_{i\ne j}  \sum_{n=0}^\infty \frac{1}{n!} \frac{\lambda_i^n-  \lambda_j^n}{ \lambda_i -  \lambda_j}
\nonumber\\
&=& \prod_{i} e^{\lambda_i} \prod_{i < j} \left(\frac{e^{\lambda_i} -  e^{\lambda_j}}{ \lambda_i -  \lambda_j}\right)^2
= e^{{\rm tr} h} \prod_{i < j} \left(\frac{e^{\lambda_i} -  e^{\lambda_j}}{ \lambda_i -  \lambda_j}\right)^2
\eea
This leads to an explicit form of the Jacobian in the functional integral.

\section{Some results in d dimensions}

First we consider the scheme where we purposefully neglect the dependence on  
(the dimensionful) $\dot F$ present inside the trace through the $\dot R_k$ term.
Iin arbitrary dimension $d$ the flow equations are
\be
\dot v=-d v +\frac{1}{2} (d-2) \phi  v'+\frac{c_d }{2} (d-1) (d-2) -c_d \frac{(d-2) f v''}{2 (d-1) f'{}^2+(d-2) f \left(1+v''\right)}
\label{vdotd}
\ee
and
\bea
\dot f \!\!\!\!&{}&= (2-d) f+\frac{1}{2} (d-2) \phi  f'-c_d\frac{\left(d^5-4 d^4-9 d^3-48 d^2+60 d+24\right)}{24 d (d-1) }+\nonumber\\
&{}&
\!\!\!\! -\frac{d c_d }{12}\frac{ \left(2 (d-1) f'{}^2+(d-2) f\right)}{2 (d-1) f'{}^2\!+\!(d-2) f \left(v''+1\right)}-
   c_d \frac{\left(2 (d-1) f'{}^2\!+\!(d-2) f\right)\! \left((d-2) f f''\!+\! 2 f'{}^2\right)}{\left(2 (d-1)  f'{}^2+(d-2) f \left(v''+1\right)\right)^2}\ ,
\label{fdotd}
\eea
where $c_d$ such that $c_d^{-1}=(4\pi)^{d/2} \Gamma(d/2+1)$.
There are three particular scaling solution of the fixed point equations.
The fixed point FP1 for constant $v$ and $f$ which is given by
\be
v_*=c_d\frac{(d-1) (d-2)}{2 d} \,, \quad f_*=-c_d \frac{\left(d^5-4 d^4-7 d^3-50 d^2+60 d+24\right)}{24 d (d-1) (d-2)}\ ,
\ee
the fixed point FP2 given by
\be
v_*=c_d\frac{(d-1) (d-2)}{2 d} \,, \quad f_*(\varphi)=f_0+\frac{1}{2(d-1)}\varphi^2\ ,
\ee
where
\be
f_0=-c_d \frac{\left(d^5-4 d^4-7 d^3-50 d^2+84 d+24\right)}{24 d (d-1) (d-2)}
\ee
and the fixed point FP3, with
\be
\label{totfp3}
v_*^1=c_d \frac{(d-2) (d-1)}{2 d} \,, \quad f_*^1(\varphi)=-\frac{(d-2) \left(d^5-4 d^4-7 d^3-50 d^2+60 d+24\right)}{8 (d-1) \left(d^5-4 d^4-7 d^3-44 d^2+72 d+24\right)} \varphi^2 \,.
\ee

Now we consider the full flow equations.
Taking into account the relations between dimensionless and dimensionful quantitites for the terms which are induced by the cutoff dependence on $F$ and $F'$,
\be
k^{2-d} \dot F=\dot f+(d-2) f-\frac{d-2}{2} \varphi  f' \,, \quad k^{1-d/2}\dot F'=\dot f'+\frac{(d-2)}{2} f'-\frac{d-2}{2} \varphi f''
\ee
the flow equations (not the fixed point equations) are more complicated and on the r.h.s. both $\dot f$, $\dot f'$ do appear.
Starting again from Eq.~(\ref{gfhess}) they are given by
\bea
\label{flowvfull}
\dot v&=&-d v+\frac{1}{2} (d-2) \varphi\,  v' +c_d \frac{(d-1) \left(d^2\!-\!d\!-\!3\right)}{d+2}+
c_d \frac{(d-2) (d+1) \left(2 \dot f-(d-2) \varphi  f'\right)}{4 \left(d+2\right) f}\\
&{}&\!\!\!\!+c_d\frac{2 \left(d^2\!-\!4\right) \!f\!+\!(d\!-\!2) \left(1\!+\!v''\right) \left(2 \dot f\!-\!6 f\!-\!(d\!-\!2) \varphi  f'\right)\!+\!
4 (d\!-\!1) f' \left(2 \dot f'\!+\!(d\!-\!1) f'\!-\!(d\!-\!2) \varphi  f''\right)}{2 (d+2) \left(2 (d-1) \left(f'\right)^2+(d-2) f \left(1+v''\right)\right)}\nonumber
\eea
\bea
 \label{flowffull}
\dot f &=&(2-d) f+\frac{1}{2} (d-2) \phi  f'-c_d\frac{d^6-2 d^5-15 d^4-46 d^3+38 d^2+96 d-24}{12 \left(d+2\right) (d-1) d}\\
&{}&\!\!\!\!-c_d\frac{\left(d^5\!-\!17 d^3\!-\!60 d^2\!+\!4 d\!+\!48\right) \left(2 \dot f\!-\!(d\!-\!2) \varphi  f'\right)}{48 (d-1) d (d+2) f} -c_d \left( (d-2) f f''+2 f'{}^2\right) \times \nonumber\\
&{}&\!\!\!\!\times \frac{(d-2) (d+2) f^2+(d-1)f'{}^2 \left((d-2) \phi  f' \!-\!2 \dot f\right)+2 (d\!-\!1) f f' \left((2-d) \varphi  f''\!+\!(d+2) f'\!+\!2
   \dot f'\right)}{(d+2) f \left(2 (d-1) \left(f'\right)^2+(d-2) f \left(v''+1\right)\right)^2} 
   \nonumber\\
   &{}&\!\!\!\!+c_d \frac{f' \left((d-2) \varphi  \left(4 (d-1) f''+(d-2) \left(v''+1\right)\right)-8 (d-1) \dot f'\right)+2 (d-2)  \left( d\, f v''-\dot f
   \left(v''+1\right)\right)}{24 \left(2 (d-1) \left(f'\right)^2+(d-2) f \left(v''+1\right)\right)}\nonumber
\eea

On setting $\dot f=\dot f'=0$ one obtains the fixed point equations.
Now one finds only two analytic solutions.
The first, which we denote again FP1, has constant $v$ and $f$ and  is given by
\be
\label{totfp1}
v_*^1=c_d \left(d-4+\frac{6}{d+2}+\frac{1}{d} \right) \,, \quad f_*^1=c_d \frac{-d^6+2 d^5+15 d^4+46 d^3-38 d^2-96 d+24}{12 (d-2) (d-1) d (d+2)}
\ee
The second is identical to (\ref{totfp3}) and is therefore called again FP3.
We note that in FP3 the second derivative of $f$ is always negative, for dimensions of interest.
The fixed point FP2 has no analogue among solutions of this equation, at least
among those for which $f$ is at most quadratic in $\phi$.
It is possible that its role is taken by a more complicated solution that we have not looked for.
Let us also stress that since the cutoff is dependent on functions of $F(\bphi)$ and $F'(\bphi)$ 
the split symmetry Ward identities will modify the flow equations and the fixed points could change again.
Their role will be the subject of further investigations.

Finally we consider the equations derived from an Hessian which is not diagonalised, 
i.e. starting from Eq.~(\ref{gfhess0}), instead from Eq.~(\ref{gfhess}) while retaining the complete $\dot F$ dependence.
For this case the flow equation for $v$ is the same while the flow equation for $f$ 
is obtained from Eq.~(\ref{flowffull}) by adding to the right hand side
\be
\Delta \dot f=
c_d\, \frac{f'}{(d+2)f}\
\frac{2(f\,\dot f'-f'\dot f)-(d-2)\left(f f'+\varphi f f''-\varphi  f'{}^2\right)}{2 (d-1) f'{}^2+(d-2) f (1+v'')} \,.
\ee
The fixed points FP1, FP3 given in (\ref{totfp1},\ref{totfp3})
are fixed points for this flow too, and again there is no analog of FP2.
This confirms the expectation that the diagonalization process is a rather mild 
change in the coarse-graining scheme.


\end{document}